\documentclass[preprint,12pt,review,sort&compress]{elsarticle}

\usepackage{inputenc}
\usepackage[T1]{fontenc}
\usepackage{graphicx}
\usepackage{tikz,tikz-inet}
\usepackage{amsmath,amssymb,amsfonts,bm}
\usepackage{textcomp}
\usepackage{subfigure}
\usepackage{booktabs,colortbl}
\usepackage[tight]{units}
\usepackage{multirow,multicol}
\usepackage{rotating}
\usepackage{hyperref}

\journal{Physica A: Statistical Mechanics and its Applications}

\begin{document}

\newdefinition{rmk}{Problem}

\begin{frontmatter}

\title{Variability Analysis of Complex Networks Measures based on	 Stochastic Distances.}

\author[cpdee,laccan]{Raquel S.\ Cabral\corref{cor1}}
\ead{raquelcabral@gmail.com}
\cortext[cor1]{Corresponding author. Tel.:+55 8232141882}
\author[laccan]{Alejandro C.\ Frery}
\ead{acfrery@gmail.com}
\author[cpdee]{Jaime A.\ Ram\'irez}
\ead{jramirez@ufmg.br}

\address[cpdee]{Graduate Program in Electrical Engineering, Federal University of Minas Gerais, Av. Ant\^onio Carlos 6627, 31270-901, Belo Horizonte, MG, Brazil}

\address[laccan]{Laborat\'orio de Computa\c c\~ao Cient\'ifica e An\'alise Num\'erica -- LaCCAN, Universidade Fe\-de\-ral de Alagoas -- Ufal, Av. Lourival Melo Mota, s/n, 57072-900, Macei\'o, AL, Brazil}

\begin{abstract}
Complex networks can model the structure and dynamics of different types of systems.
It has been  shown that they are characterized by a set of measures. 
In this work, we evaluate the variability of complex networks measures face to perturbations and, for this purpose, we impose controlled perturbations and quantify their effect. 
We analyze theoretical models (\textit{random}, \textit{small-world} and \textit{scale-free}) and real networks (\textit{a collaboration network} and \textit{a metabolic networks}) along with the \textit{shortest path length}, \textit{vertex degree}, \textit{local cluster coefficient} and \textit{betweenness centrality} measures.

In such analysis, we propose the use of three stochastic quantifiers: the Kullback-Leibler divergence and the Jensen-Shannon and Hellinger distances.
The sensitivity of these measures was analyzed with respect to the following perturbations: edge addition, edge removal, edge rewiring and node removal, all of them applied at different intensities. 
The results reveal that the evaluated measures are influenced by these perturbations. 
Additionally, hypotheses tests were performed to verify the behavior of the degree distribution to identify the intensity of the perturbations that leads to break this property.

\end{abstract}


\end{frontmatter}

\section{Introdution} \label{sec:introdution}

Complex networks are systems whose structure is irregular, complex and dynamically evolving in time~\citep{Boccaletti2006}. 
In recent years, a number of measures have been developed to quantify the structure and behavior of such systems, which provide a framework that allows its characterization, analysis and modeling, reflecting different features of the network such as, connectivity, centrality, cycles, distances, among others.

The choice of an appropriate measure for the characterization of a network is performed by evaluating its behavior, and depends mainly on three factors: (i)~data availability, (ii)~storage capacity and processing and (iii)~interest in characterizing the behavior of the measures. 
In this procedure, the network is mapped into a feature vector~\citep{Costa2007}; however, in many cases the mapping is not complete and does not accurately describe the network's real structure.
In such cases, it is important to evaluate the performance of the measures when unexpected changes occur in the networks.
For instance, \textit{what is the behavior of the measures if the network loses links or nodes?} or, \textit{do these changes break the properties used to describe the network structure?} 
To address such problems, it is necessary to compare different states of the network.
In this work, we investigate the use of methods from the Information Theory, in particular the concept of Stochastic Quantifiers, as means to quantify the changes.

There are a few works that explore the use of stochastic measures to analyze the behavior of complex networks.
\citet{Wang2006} employed the concept of Entropy of the degree distribution to provide an amount of the network's heterogeneity, since that measure quantifies the diversity of the link distribution.
They also studied the robustness of scale-free networks using the Entropy. 

A common practice to address the questions raised previously is to use samples of the network, instead of considering it entirely.
This saves memory and processing time.
\citet{Boas2010} employed the Kullback-Leibler divergence to compare different states of the network and assessed the appropriateness of using network samples. 
However, this divergence can not be considered a distance since it is not commutative.
We extend those results in two manners, namely, (i)~we analyze the behavior of four measures applied to different theoretical network models and real networks, and (ii)~we compare the sensitivity of three stochastic quantifiers with respect to several perturbations of the networks, including node removal.\label{ComentarioAoArtigoDeBoas}

The study of the evolution of complex networks is an important issue due to their dynamic nature.
Among the changes a network may be subjected to, we will consider node removal, edge addition, edge removal, and edge rewiring.
Such perturbations describe common changes in practical situations as, for instance, the death of a node, the creation and the deletion of a connection, and both operations at once.\label{ExplicaPerturbacoes}

\citet{Carpi2011} proposed a new quantifier based on Information Theory for the analysis of the evolution of small-world dynamic networks.
The quantifier, a statistical complexity measure, is used to compute changes in the topological randomness for degree distribution of the network. 
It is  obtained by the product of the normalized Shannon entropy and the normalized Jensen-Shannon distance.
This quantifier requires the use of a probability distribution as a reference to compute the Jensen-Shannon distance. The authors used three reference distributions: Poisson, uniform and the distribution corresponding to the regular lattice.

The analysis of perturbations in networks has also been studied with two practical purposes in mind: their vulnerability to attacks and the identification of elements whose failure lead to a breakdown.
The vulnerability is associated with the decrease of network performance when structural changes occur; these can be caused by the random or directed removal of vertices, termed failure and attack, respectively.
Measures related to this property are commonly defined in terms of the shortest path length and the size of the connected components of the graph.
The main idea is to intentionally apply a sequence of failures (or attacks) to the network, and to observe its behavior~\cite{Albert2000,Zhang2012,Ghamry2012,Mishkovski2011,Manka-Krason2010}.

In this context, many real networks have been studied.
\citet{Pu2012} studied the behavior of network controllability under vulnerability for networks of different topologies under two different kinds of attacks, including the power grid networks. 
\citet{Jeong2001} evaluated the diameter behavior in protein networks by removal the most connected protein, and \citet{Kaiser2004} evaluated the robustness toward edge elimination of metabolic networks showing that intercluster connections represent the most vulnerable edges in these networks.

The breakdown phenomenon in networks refers to a type of cascade process, where the failure of a single or a few nodes may result in the collapse of their functionality.
In networks with power distribution, for instance, the failure of a node requires that its load is redistributed to other nodes, causing a network overload and possibly other faults~\cite{Wang2012,Huang2008}.

\citet{Cabral2013} presented the analysis of communication strategies in wireless sensor networks by means of analyzing the variation of the shortest path length measure in flooding, random, small-world, and scale-free networks.
The variation of this measure was analyzed with respect to the insertion and removal of nodes in flooding; and with respect to insertion, removal and rewiring of links in the strategy based in complex networks. 
Stochastic quantifiers, namely the normalized Kullback-Leibler divergence and Hellinger distance were used to quantify the variation of the shortest paths. 

The goal of this work is to analyze the behavior of measures face to network perturbations.
We propose the use of quantifiers based on information theoretic tools to perform this analysis.  
The complex network measures evaluated were shortest paths length, vertex degree, local cluster coefficient and betweenness centrality.
The first one is calculated for each pair of nodes and the others are calculated for each network node. 
We compare the results, which are vector-valued measures, using of stochastic divergences between discrete probability distributions.
In addition, we propose the use of three quantifiers: the Kullback-Leibler divergence, the Jensen-Shannon and Hellinger distances.

The proposed methodology allows the comparison of the behavior of different quantifiers and the identification of the intensity of the perturbations that leads to significant changes of their properties.   
The results reveal that all the evaluated measures are influenced by the perturbations considered, but to different extent.

The paper is organized as follows.
Section~\ref{sec:problem_definition} defines the problem that is investigated in this work. 
Section~\ref{sec:methods} presents the definitions and methods used in this paper.
Section~\ref{sec:results} presents the results of measures behavior and quantifiers performance.
Finally, conclusions are presented in Section~\ref{sec:conclusion}.
\section{Problem definition} \label{sec:problem_definition}

A complex network can be described as an undirected graph $\mathcal G = (\mathcal V,\mathcal E)$, where $\mathcal V = \{v_1, v_2, \ldots, v_N\}$ is the set of vertices (or nodes), and  $\mathcal E = \{e_1,e_2,\ldots,e_M \mid e_m = (v_i, v_j): 1 \leq i, j \leq N, v_i \neq v_j\}$ is the set of edges (or links) between nodes~\citep{Bollobas1998}. 
A graph $\mathcal G$ is represented by an adjacency matrix $\mathcal{A}_{N \times N}$ with elements $a_{ij}$, where $i,j = 1,\ldots,N$.
The element $a_{ij}$ is equal to one when the link between $v_i$ and $v_j$ exists, and zero otherwise.

In this way, the problem addressed in this work can be stated as following:

\begin{rmk}
Let $\mathcal G$ be a class of complex networks, $\mathcal F = \{f_1,f_2,\ldots,f_F\}$ a vector of features of $\mathcal G$, and $\Pi=\{\pi_1,\pi_2,\ldots,\pi_P\}$ a set of graph perturbations. 
We are interested in the behavior of $\mathcal{F}$ applied to $\mathcal{G}$ before $\Pi$.
\end{rmk}

In other words, we want to know how the perturbations under assessment change the intrinsic characteristics of the considered class of graphs

To do this analysis we take elements of $\mathcal G $, say $g_1,g_2,\ldots,g_Z$, and subject each to a series of perturbations. 
For example, $g_j$ is transformed into $\eta_{1,j} = \pi_1(g_j), \eta_{2,j} = \pi_2(g_j),\ldots, \eta_{i,j} = \pi_i(g_j)$.
In this way, we have $\Pi \colon \mathcal G = ( \mathcal V,\mathcal E) \rightarrow \mathcal G' = (\mathcal V',\mathcal E')$, where $\mathcal V'$ and $\mathcal E'$ are the new sets of vertices and edges, defined according to the type of perturbation $\Pi$ applied to $\mathcal G$.

In complex networks, four types of perturbations are very common and were applied in this methodology: 
\begin{itemize}
\item Edge addition $\pi_{ae}$: an edge $(v_i,v_j) \not\in \mathcal{E}$ is added on $\mathcal G$, setting $\mathcal E' = \mathcal{E} \cup (v_i,v_j)$, i.e., $\pi_{1}\colon\mathcal E = \{e_1,e_2,\ldots,e_{M} \} \rightarrow \mathcal E' = \{e_1,e_2,\ldots,e_{m}, e_{M+1} \}$.
\item Edge removal $\pi_{re}$: an edge $(v_i,v_j) \in \mathcal E$ is removed from $\mathcal G$, then
$\mathcal E' = \mathcal E \setminus (v_i,v_j)$ i.e., $\pi_{2}:\mathcal E = \{e_1,\ldots,e_{m-1},e_m, e_{m+1},\ldots,e_M \} \rightarrow \mathcal E' = \{e_1,e_{m-1},e_{m+1},\ldots,e_{M-1}\}$.
In this case the graph may become disconnected. 	  
\item Edge rewiring $\pi_{we}$: two edges $((v_i,v_j),(v_k,v_l)) \in \mathcal E$ are substituted  for $((v_i,v_l),(v_j,v_k)) \not\in \mathcal E$ on $\mathcal G$, then $\mathcal E' = \mathcal E \setminus ((v_i,v_j),(v_k,v_l)) \cup ((v_i,v_l),(v_j,v_k))$, i.e., $\pi_{3}\colon\mathcal E = \{e_1,e_2, \ldots,e_{M} \} \rightarrow \mathcal E' = \{e'_1,e'_{i-1}, \ldots, e'_{M}\}$. 
\item Node removal $\pi_{rv}$: a vertex $v_j \in \mathcal{V}$ is removed from $\mathcal G$, then $\mathcal V' = \mathcal V \setminus v_j$ and $\mathcal E' = \mathcal E \setminus (v_j, v_i)$, where $v_i$ are the vertices linked to $v_j$, i.e., $\pi_4 \colon \mathcal V = \{v_1, \ldots, v_{n-1}, v_{n}, v_{n+1}, \ldots, v_{N} \} \rightarrow \mathcal V' = \{v_1, \ldots, v_{n-1}, v_{n+1}, \ldots,v_{N-1} \}$ and $\pi_4 \colon \mathcal E = \{e_1, e_2, \ldots, e_{M} \} \rightarrow \mathcal E' = \{e_1, e_2, \ldots, e_{M-x}\}$, where $x$ is the number of edges which were linked to $v_j$ before the removal.  
\end{itemize}
		
We then study sample quantities (measures) of the perturbed samples and use the stochastic quantifiers to compare them.
\section{Methods} \label{sec:methods}

In this section, we focus on the description of the complex network measures, models and stochastic measures used to evaluate the effects of perturbations on networks.

\begin{figure}[htb]
 \centering
\begin{tikzpicture}[node distance = 1cm]
\node (n0){$g_j$};
\node[below =of n0](n1){$f_k(g_j)$ };
\node[right =4cm of n0](n2){$\eta_{ij}$};
\node[below =of n2](n3){$f_k(\eta_{ij})$};
\node[below left=2cm and 0.4cm of n2](n6){$\mathcal{H}[f_k(g_j)]$,$\mathcal{H}[f_k(\eta_{ij})]$};
\node[below =of n6](n7) {$\mathcal D$};

\path [draw,thin,->] (n0) -- (n1) node[midway,left]{$f_k \in \mathcal{F}$}; 
\path [draw,thin,->] (n0) -- (n2) node[midway,sloped,above]{$\pi_i \in \Pi$};
\path [draw,thin,->] (n2) -- (n3) node[midway,right]{$f_k \in \mathcal{F}$}; 
\path [draw,thin,->] (n1) -- (n6);
\path [draw,thin,->] (n3) -- (n6);
\path [draw,thin,->] (n6) -- (n7);
\end{tikzpicture}
\caption{Steps of the variability analysis of measures with stochastic distances.}
\label{fig:etapasAnalise}
\end{figure}
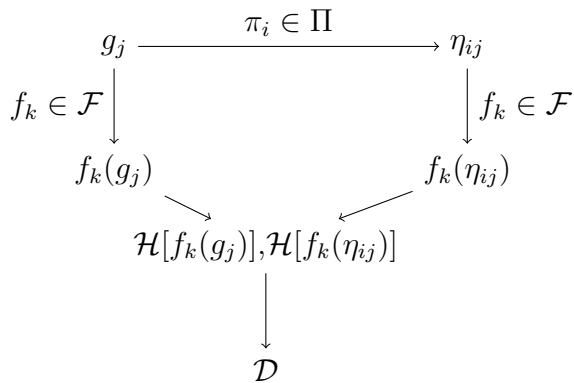

Figure~\ref{fig:etapasAnalise} shows the steps to perform the analysis.
Given the initial network $g_j$, first we calculate the measures $\mathcal{F}$ and obtain, for instance, $f_k(g_j)$. 
We then apply the perturbation $\pi_i$ to $g_j$, produce the network $\eta_{ij} =\pi_i(g_j)$ and calculate the measures $\mathcal{F}$ for $\eta_{ij}$ and obtain, for instance, $f_k(\eta_{ij})$.
Then $f_k(g_j)$ and $f_k(\eta_{ij})$ are transformed into probability functions (histograms) $\mathcal{H}[f_k(g_j)]$ and $\mathcal{H}[f_k(\eta_{ij})]$ in order to compare them with the stochastic quantifiers $\mathcal{D}$.
Stochastic quantifiers have been shown to exhibit good discriminatory properties in a number of problems~\cite{Boas2010,Carpi2011}.

\subsection{Measures}\label{subsec:measures}

The structural analysis or characterization of a network can be made with a visual inspection of their corresponding graphs if we have a small number of nodes.
However, with the emergence of networks with thousands, millions even billions of nodes, like the Internet, this task became impossible.
The solution was to map the networks in terms of topological measures. 
So, we have a feature vector $\mathcal{F} = \{f_1,f_2,\ldots,f_F\}$ for the network $\mathcal G$, i.e., a vector of related measures such as, vertex degree, clustering coefficient, the network diameter, and so on~\citep{Costa2007}. 

The problem is: \textit{what is the behavior of the measures if the network loses links or nodes?} or, \textit{these changes lead to break the properties used to describe the network structure?} 
In this work, we evaluate the behavior of some measures associated to specific perturbations. 

There are plenty of measures that characterize different networks. This section presents the definition of the measures evaluated in this work.

The measures chosen for the analysis are among the most used to characterize a large number of networks. 
They are also used as the basis for defining of other measures, moreover, they reflect important network features.
They are defined as follows:
\begin{description}
\item[Shortest path] $\ell_{(i,j)}$, or geodesic path, is any path that connects the vertices $i$ and $j$ with minimal distance. 
This is a measure of the efficiency of information or mass transport on a network.
\item [Vertex degree] $\kappa_i$ of a node $i$ in a network is the number of edges incident to it, and is defined in terms of the adjacency matrix $\mathcal{A}_{N \times N}$ as:
\begin{eqnarray}
\kappa_i &=& \sum_{j \in \mathcal{V}} a_{ij}.
\label{eq:degree}
\end{eqnarray}
The degree distribution $p(\kappa)$ of a graph $\mathcal G$ is, probably, the most wide\-spread topological characterization of graphs.
It is defined as the probability that a node chosen uniformly at random has degree $\kappa$, i.e., the fraction of nodes in the graph with degree $\kappa$~\citep{Boccaletti2006}.
\item [Local cluster coefficient] $C_i$, or transitivity, is a measure that characterizes grouping in complex networks. 
First introduced to unweighted graphs by~\citet{Newman1999}, this measure can be defined as the number of the edges (denoted by $ne_i$) that connects neighbors of node $i$:
\begin{eqnarray}
C_i &=& \frac{2ne_i}{\kappa_i(\kappa_i -1)},
 \label{eq:Ci}
\end{eqnarray}
where  $\kappa_i(\kappa_i-1)/2$ is the maximum number of edges in the sub-graph of neighbors of $i$.
\item[Betweenness centrality] $B_u$ is a measure of a vertex's centrality in a network.
It is equal to the number of shortest paths of the network that pass through that vertex $v_u$, defined as: 
\begin{eqnarray}
B_u &=& \sum_{i,j \in \mathcal V} \frac{\sigma(i,u,j)}{\sigma(i,j)},
 \label{eq:Bu}
\end{eqnarray}
where  $\sigma(i,u,j)$ is the number of shortest paths between $v_i$ and $v_j$ that pass through $v_u$, $\sigma(i,j)$ is the total number of shortest paths between $v_i$ and $v_j$.
\end{description}

The first three measures are used to characterize the theoretical network models and the latter identifies important components in the network.

\subsection{Complex networks}\label{subsec:models}

We employ both theoretical models and real networks to study the effect of perturbations in complex networks.
The theoretical models are widely used to describe real networks, specifically, Erd\H{o}s R\'{e}nyi's random model, Watts-Strogatz' small-world model, Barab\'{a}si-Albert's scale-free model.
The real-world networks are the metabolic and collaboration networks. 

The main characteristics of these networks are as follows.
\begin{description}
\item[Random model] $\mathcal{G}_{ER}$, in this model the probability to connect each pair of nodes is the same.
The degree distribution of this model is a Poisson distribution:
\begin{equation}  
p(\kappa) = \frac{e^{-c} c^{\kappa}}{\kappa!}, 
\label{eq:pk_er} 
\end{equation}
where $c$ is the mean degree of network.
There are two ways to build a random graph~\citep{Erdos1960}: $N$ nodes and exactly $M$ links, and $N$ nodes and the probability $0<p_c<1$ to connect each pair of nodes.
In the first description, $M$ links are uniformly distributed among the $N(N-1)/2$ possibilities.
In the second description, which is the one we adopted here, we start with a totally disconnected graph and then connect each pair of nodes with probability $p_c$, so the probability of observing $0\leq \kappa \leq N-1$ connections in each node follows a Binomial distribution with $N-1$ trials and probability of success $p_c$, leading to $c = (N-1)p_c$.
\item[Small-world model] $\mathcal{G}_{WS}$, presents a high clustering coefficient and a small shortest path length, so it is a  model  between regular and random networks.
The degree distribution of this model is:
\begin{equation}
p(\kappa) =
\sum_{i = 1}^{\min\{\kappa - c/2,c/2\}} \binom{c/2}{i} (1 - p_r)^{i} p_{r}^{c/2 - i} \frac{(p_r c/2)^{\kappa-c/2-i}}{(\kappa - c/2 - i)!}e^{-p_r c/2}, 
\label{eq:pk_ws} 
\end{equation}
where $\kappa \geq c/2$, $c$ is the is the mean degree of network and $p_r$ is a probability of rewiring edges.
If $\kappa < c/2$ then $p(\kappa) = 0$.

There are several ways to build such network; in this work we use the Watts-Strogatz model. 
This model starts with a circular regular topology with $N$ nodes, each one connected to the $c/2$ nearest neighbors in each direction (right or left in circular topology). 
Then, each link is randomly ``rewired'' with probability $p_{r}$~\citep{Watts1998}.
\item[Scale-free model] $\mathcal{G}_{BA}$,  it displays a power law degree distribution 
\begin{equation}
p(\kappa) \sim \kappa^{-\lambda}, \kappa > 0 \ \text{and} \ 2 < \lambda < 3. \label{eq:pk_sf}
\end{equation}
The main feature of this topology is the presence of a few nodes with high degree, often called ``hubs''. 
To generate this topology we use the Barab\'{a}si-Albert scale-free model that starts with a small number of nodes $n_0 \geq 2$ with $m_0$ links, and in the next step a new node with $m$ links is added to the network (with $m \leq  n_0$).
The probability connection is linearly proportional to the node degree~\citep{Barabasi1999}.

\item[Metabolic networks] Metabolic and physical processes that determine the physiological and biochemical properties of a cell can be described by $\mathcal{G}_{ME}$.
The structure of these networks involves hundreds or thousands of components, for this reason they have been widely studied as complex networks. 
The metabolic organization is not identical for all organisms, but has small-world properties and its connectivity follows a power law, as in scale-free networks~\cite{Jeong2000}.
In this case, substrates represent the nodes of a metabolic network, while links represent the chemical reactions the substrates participate in. 
An undirected graph is obtained by linking all in-coming substrates (educts) of a reaction to all its outgoing substrates (products).

\item[Scientific collaboration networks] $\mathcal{G}_{SC}$ are social networks where the nodes are the scientists and the edges represent their coauthors. 
According to~\cite{Newman2001a} these networks present small-world properties and are highly-clustered (with clustering coefficient between 0.30 and 0.40). 
\end{description}

Next section examines the stochastic measures used in this work, that is, the use of tools derived from information theory, specifically Kullback-Leibler divergence, and Jensen-Shannon and Hellinger distances.

\subsection{Stochastic quantifiers} \label{subsec:stochastic_distances}

In probability theory and statistics, metrics, divergences and distances are popular measures for comparing different probability distributions.
In particular, Information Theory provides tools known as divergence measures based on the concept of entropy to statistically discriminate stochastic distributions.
The normalized Kullback-Leibler divergence, the Jensen-Shannon and Hellinger distances are three quantifiers suitable for describing the difference between distributions~\citep{Nascimento2010}.

We perform the analysis of the variability of complex networks measures $\mathcal{F}$ and compare different states of the network with respect to the set of perturbations $\Pi$. 

Consider the discrete random variables $X$ and $Y$ defined on the same sample space $\Omega = \{\xi_1, \xi_2, \dots, \xi_n \}$.
The distributions are characterized by their probability functions $p, q \colon \Omega \to [0,1]$, where $p(\xi_i) = \Pr(X=\xi_i)$ and $q(\xi_i) = \Pr(Y=\xi_i)$.
A metric $\mathcal{D}$ between these two distributions is a quantifier obeying:  
\begin{enumerate}
\item\label{prop:reflex} $\mathcal{D}(p,p) = 0$, reflexivity; 
\item\label{prop:nonnegat} $\mathcal{D}(p,q) \geqslant 0$, non-negativity; 
\item\label{prop:commute} $\mathcal{D}(p,q) = \mathcal{D}(q,p)$, commutativity; 
\item\label{prop:triangle} $\mathcal{D}(p,q) \leqslant \mathcal{D}(p,r) + \mathcal{D}(r,q)$, triangle inequality for any other probability function $r$ defined on the same probability space.
\end{enumerate}
A distance is not required to satisfy property~\ref{prop:triangle} and a divergence is only required to satisfy properties~\ref{prop:reflex} and~\ref{prop:nonnegat}~\citep{Cha2002}.

Assuming $q(\xi)>0$ for every event $\xi\in\Omega$, the Kullback-Leibler divergence is defined as:
\begin{equation}
 D_{KL}(p,q) = \sum_{\xi\in\Omega} p(\xi) \log \frac{p(\xi)}{q(\xi)} .\label{eq:entropiaRelativa}
\end{equation}

The Shannon Entropy $\mathcal S(p)$ of the distribution $p$ is given by 
$$\mathcal S (p) = - \sum_{\xi \in \Omega} p(\xi) \log p(\xi),$$ 
with the convention $0 \ln 0 = 0$. 
The Jensen-Shannon distance is defined as
\begin{eqnarray}
 \mathcal D_{JS}(p,q) =  \frac{1}{2}\big(\mathcal S(p) + \mathcal S(q)\big) - \mathcal S \Big(\frac{p + q}{2} \Big),\label{eq:shannon_D_JS}
\end{eqnarray}

The Hellinger distance does not impose positivity on the probabilities; it is defined as
\begin{equation}
 \mathcal{D}_H^2(p,q) = \frac{1}{2}
 \sum_{\xi \in \Omega}\Big( 
 \sqrt{p(\xi)} - \sqrt{q(\xi)}
 \Big)^2 
\label{eq:hellinger}
\end{equation}

In order to make the Kullback-Leibler divergence (an unbounded positive quantity) and the Hellinger distance (which is confined to the $[0,1]$ interval) comparable, in the remainder of this work we will use the normalized Kullback-Leibler distance defined as ${\mathcal D}_{KL}(p,q)=1-\exp\{-D_{KL}(p,q)\}$.
\section{Results} \label{sec:results}

This section presents a simulation study which evaluates the behavior of measures from complex network with respect to perturbations, i.e., we are interested in the behavior of $\mathcal{F}$ applied to $\mathcal{G}$ associated to $\Pi$. We take elements of $\mathcal G$ and subject each to a series of perturbations comparing different states of the networks with different quantifiers. 
The study presented here shows the application of the methodology outlined in Figure~\ref{fig:etapasAnalise}, limited to the following scope: 
the theoretical models examined are the \textit{random}, \textit{small-world} and \textit{scale-free}, and real networks examined are \textit{collaboration networks} and \textit{metabolic networks}.
The measures considered are the shortest paths length ($\ell_{(i,j)}$), vertex degree ($\kappa_i$), local cluster coefficient ($C_i$) and betweenness centrality ($B_i$), so $\mathcal{F} = \{\ell_{(i,j)}, \kappa_i, C_i, B_i \}$. 
The perturbations applied are the edge addition ($\pi_{ae}$), edge removal ($\pi_{re}$), edge rewiring ($\pi_{we}$) and node removal ($\pi_{rv}$), so $\Pi=\{\pi_{ae},\pi_{re},\pi_{we},\pi_{rv}\}$.
In addition, to evaluate the behavior of $\mathcal{F}$, we use as quantifiers the Kullback-Leibler divergence ($\mathcal{D}_{KL}$) and two distances: Jensen-Shannon ($\mathcal{D}_{JS}$) and Hellinger ($\mathcal{D}_{H}$).

The simulation assumptions and parameters were:
\begin{description}
\item[Theoretical model parameters:] networks with $N = \{ 1000,5000,10000 \}$ vertices and average degree equal to $c = 6$ for the three models.
The probability of connection ($p_c$) in the random model was $0.06$, $0.012$ and $0.0006$ to $1000$, $5000$ and $10000$ nodes, respectively.
The nearest neighbors and probability rewiring in small-world models was three and $p_r = 0.3$, respectively.
The number of links added in each step in the scale-free model was $m = 1$.

\item[Real Network data:] two real data were evaluated, namely:
\begin{itemize}
\item The WIT database~\citep{Overbeek2000} provides descriptions for the metabolic networks of forty-three different organisms based on data for six arch\ae a, thirty-two bacteria and five eukaryotes organisms.
We analyzed the metabolic network of Escherichia coli, which has $778$ nodes.
The data were downloaded from Center for Complex Networks Research\footnote{\url{http://www3.nd.edu/~networks/resources.htm}}, the graphs were those presented in~\citet{Jeong2000}.

\item In the Scientific Collaboration networks we examined the collaboration network of scientists posting preprints on the astrophysics archive at \url{www.arxiv.org} (\textit{astro-ph}).
This network has $16706$ nodes and was created by \citeauthor{Newman2001a}~\citep{Newman2001a,Newman2006}\footnote{The data set is available in \url{http://www-personal.umich.edu/~mejn/netdata/}}. 


\end{itemize}

Twenty perturbations of each kind were applied to each network

\item [Perturbations:] we performed four types of perturbations, namely: 
	\begin{description}
	\item{\textit{edge removal}}, in which edges are selected at random uniformly and removed from the network; 
	\item{\textit{edge addition}}, in which two unconnected vertices are selected at random and a new edge is established between them; 
	\item{\textit{edge rewiring}}, in which two pairs of connected vertices are randomly chosen and their connections exchanged; and
	\item{\textit{node removal}}, in which vertices are selected at random and removed from the network together with its edges.
	\end{description}
	
Perturbations were applied according to the network model. 
We used edge additon, edge removal, edge rewiring and node removal in theoretical models, and edge addition, edge removal and node removal in real networks.
Following \citet{Boas2010}, these perturbations were performed in $\{ 1\%,2\%,\ldots,10\% \}$ of the total number of edges (edge perturbations) and of nodes (node removal).
 
\item [Histograms:] for each network (theoretical and real), the normalized histogram $\mathcal{H}$, also known as histogram of proportions, was obtained with $200$ bins of equal width.
The $\mathcal{D}_{KL}$ diverges for $q(\xi) = 0$ and $p(\xi) \neq 0$, as defined in equation~\eqref{eq:entropiaRelativa}. For the calculations, a small positive constant $\delta = 0.001$ was added to each bin and then the histogram is normalized to add one~\citep{Cabella2008}. 
The original histogram is used to compute the Hellinger distance once it does not impose the positivity restriction on the probabilities.

\item [Simulation parameters:] for each model network (theoretical and real) we generated 10 different networks and for each network 20 different perturbations were made.  
Each replication refers to a state of the network $g_j$ after a perturbation $\pi_i$, denoted by $\eta_{ij}=\pi_i(g_j)$. 
In this way we are able to present the mean results with symmetrical asymptotic confidence intervals at the $95\%$ significance level.

\item [Computational resources:]  we performed our evaluation using the \texttt R platform~\citep{Rmanual}, on an Intel(R) Core(TM) i5 CPU 760 \unit[2.80]{GHz}  with \unit[8]{GB} RAM, running Ubuntu 12.04 (\unit[64]{bits}).
The \texttt{igraph} library was used to generate and modify the graphs~\citep{igraph}.
\end{description}


\subsection{Theoretical models}

\begin{sidewaysfigure}[htb]
\begin{center}
	\subfigure[Random model]{
	\includegraphics[width=.45\textwidth]{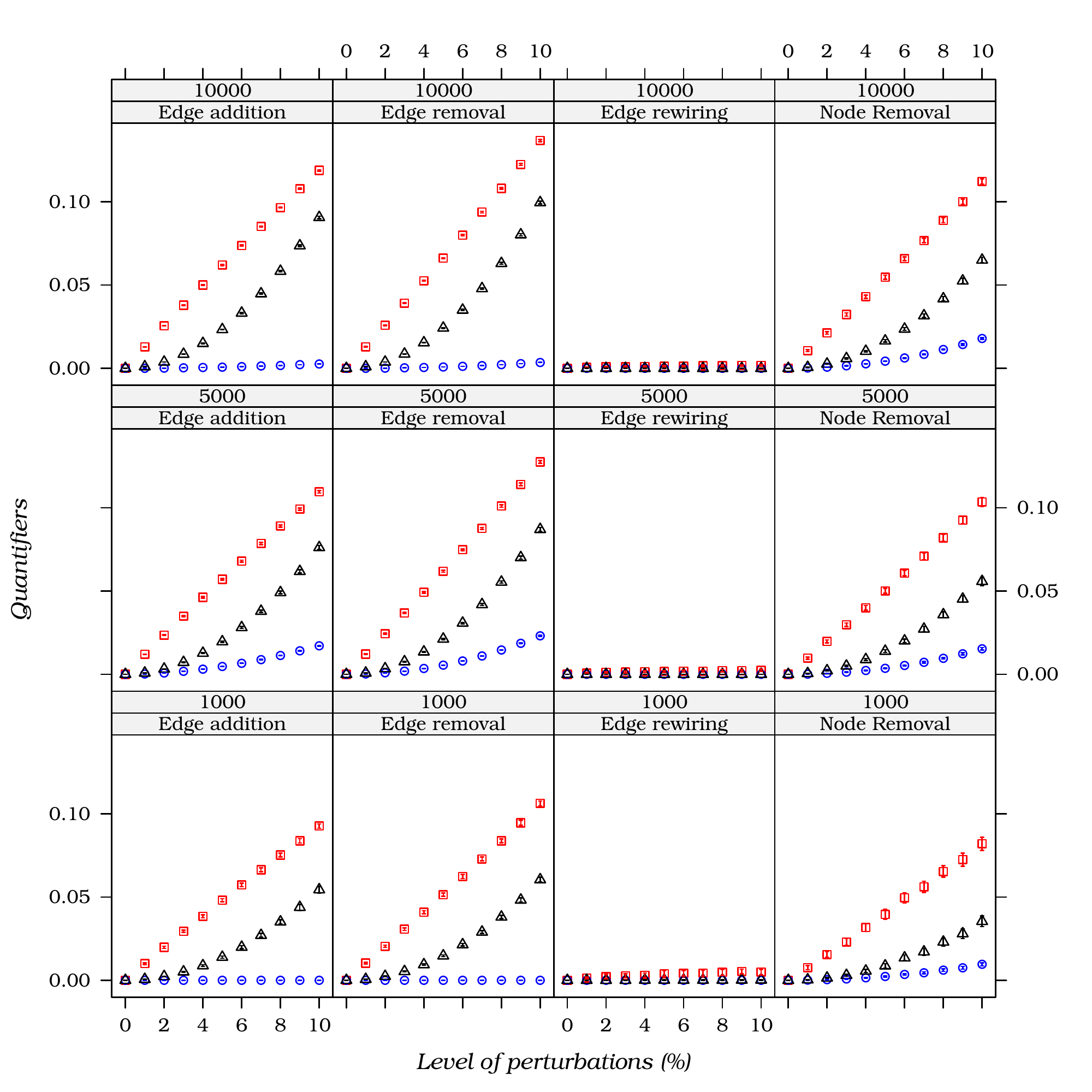}\label{fig:D_ER_l}}
	\subfigure[Small-world model]{
	\includegraphics[width=.45\textwidth]{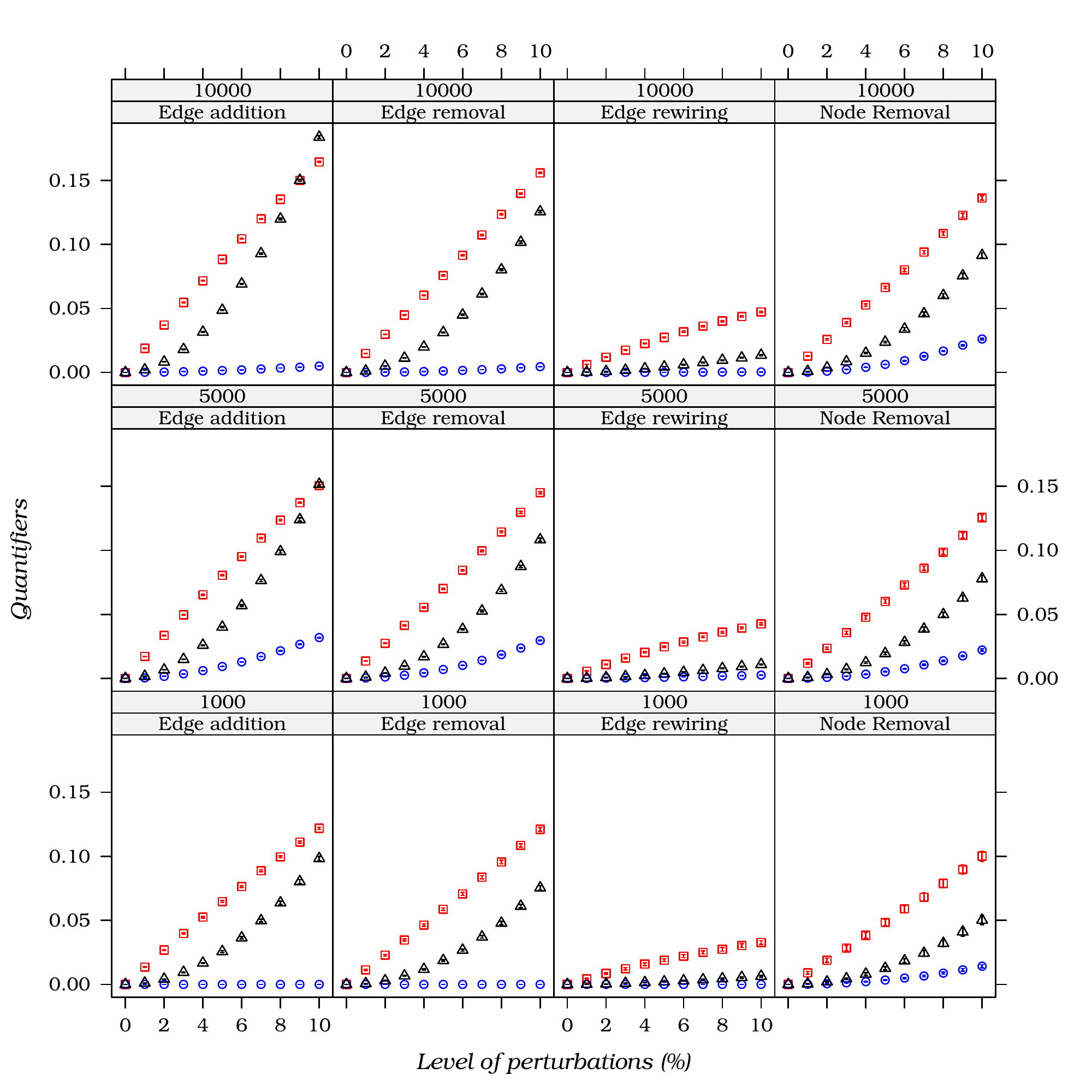}\label{fig:D_WS_l}}
	\subfigure[Scale-free model]{
	\includegraphics[width=.45\textwidth]{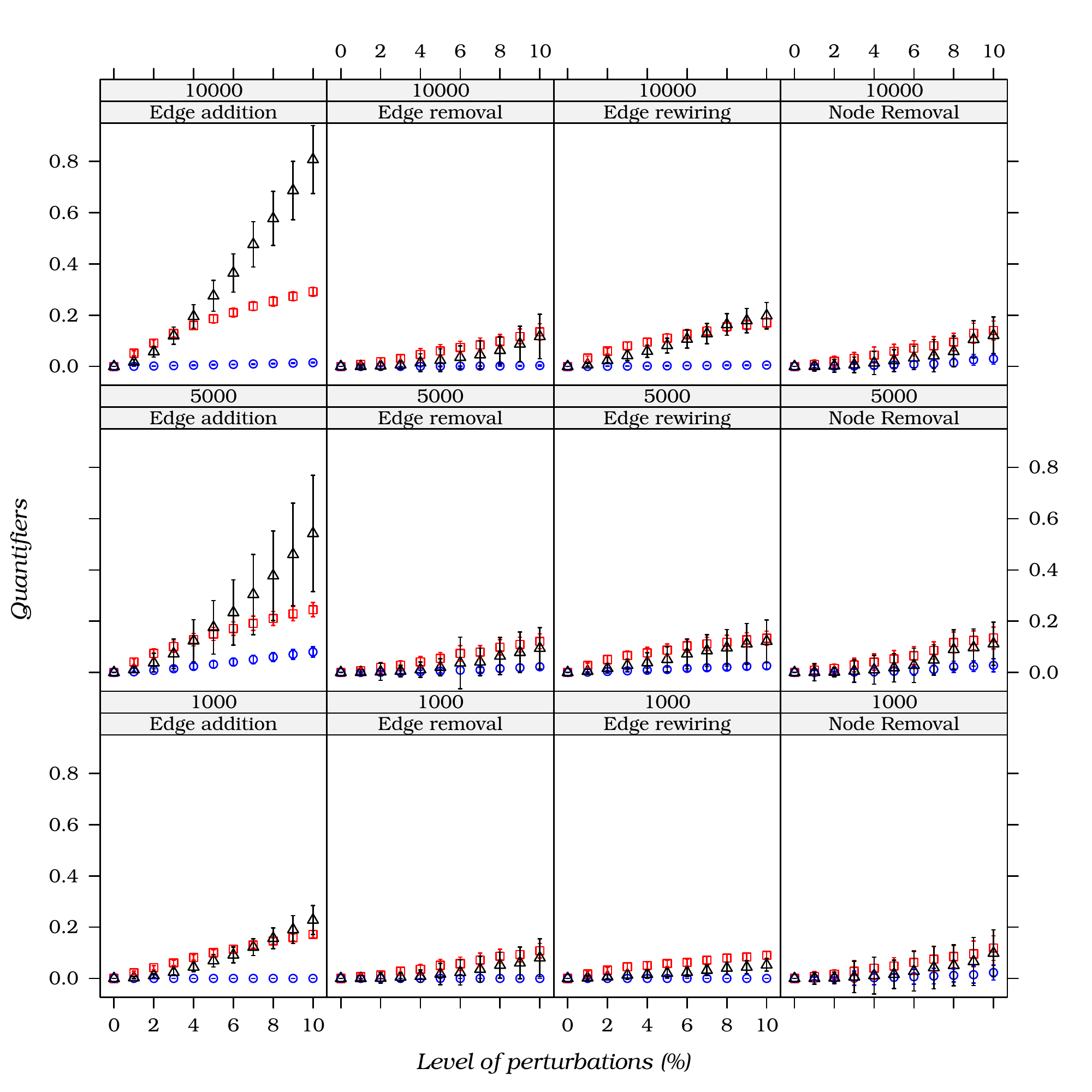}\label{fig:D_BA_l}}
	\caption{Hellinger distance ($\square$), Jensen-Shannon distance ($\circ$) and Kullback-Leibler divergence ($\triangle$) quantifiers for the shortest paths $\ell_{(i,j)}$.}
	\label{fig:D_l}
\end{center}
\end{sidewaysfigure}

\begin{sidewaysfigure}[htb]
\begin{center}
	\subfigure[Random model]{
	\includegraphics[width=.45\textwidth]{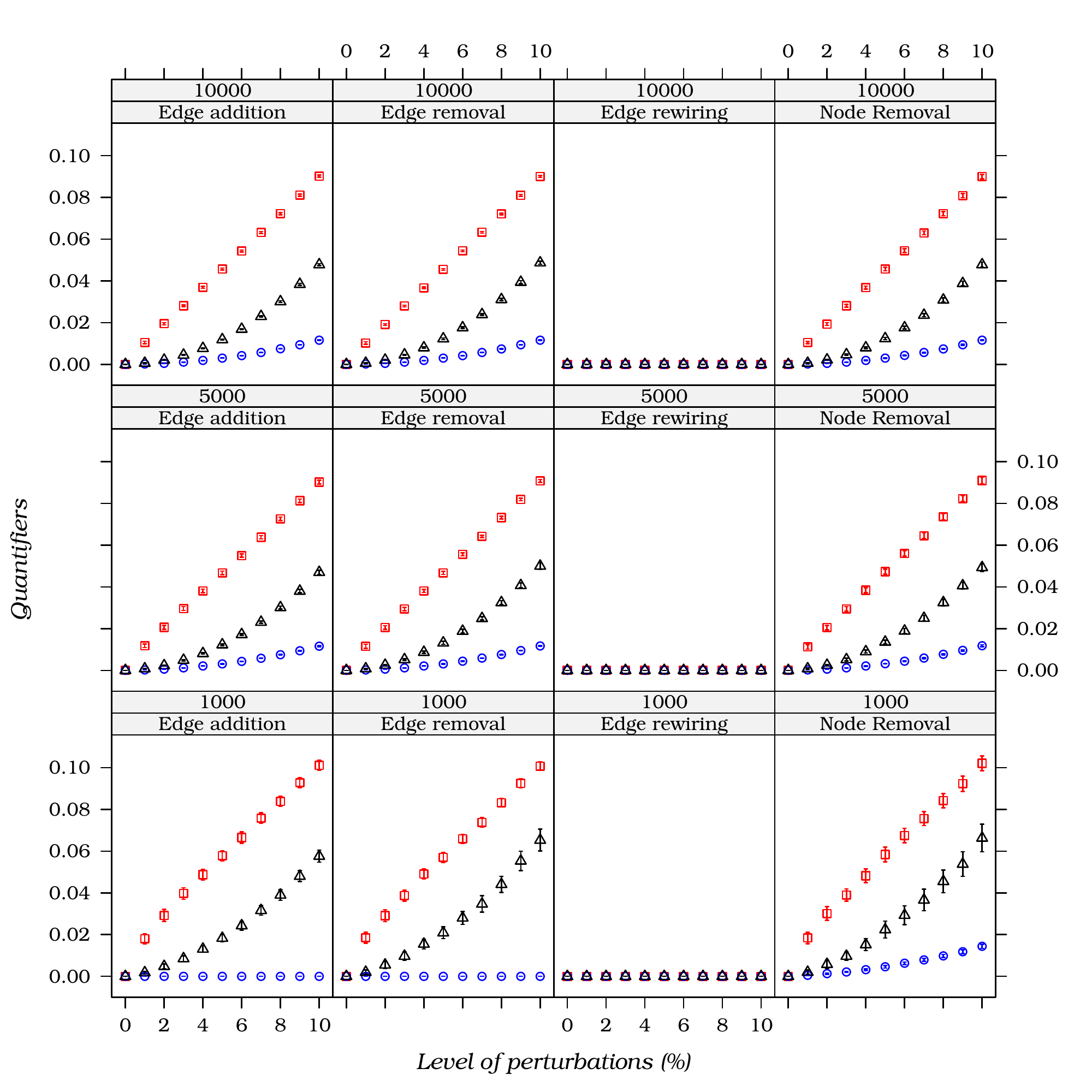}\label{fig:D_ER_k}}
	\subfigure[Small-world Model]{
	\includegraphics[width=.45\textwidth]{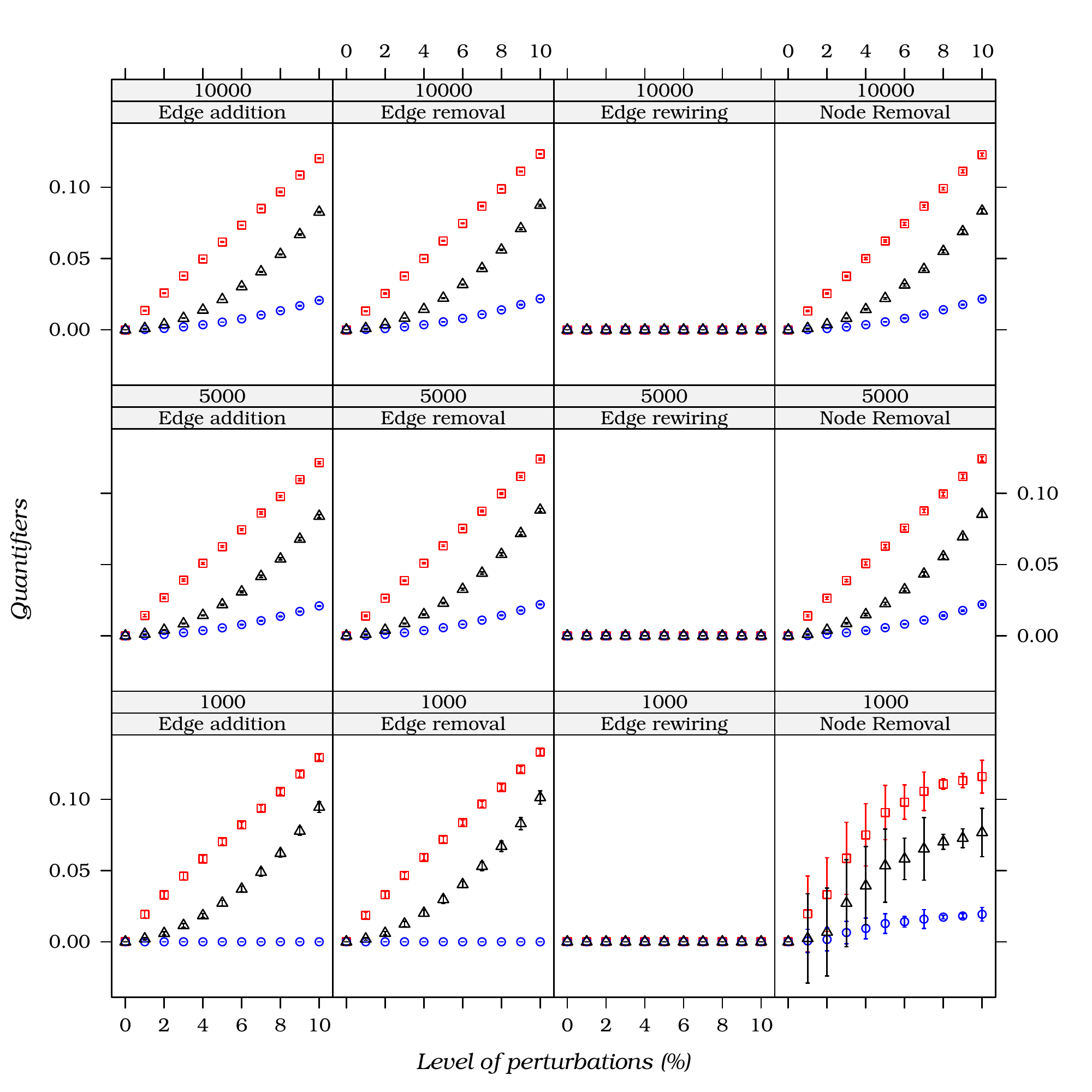}\label{fig:D_WS_k}}
	\subfigure[Scale-free model]{
	\includegraphics[width=.45\textwidth]{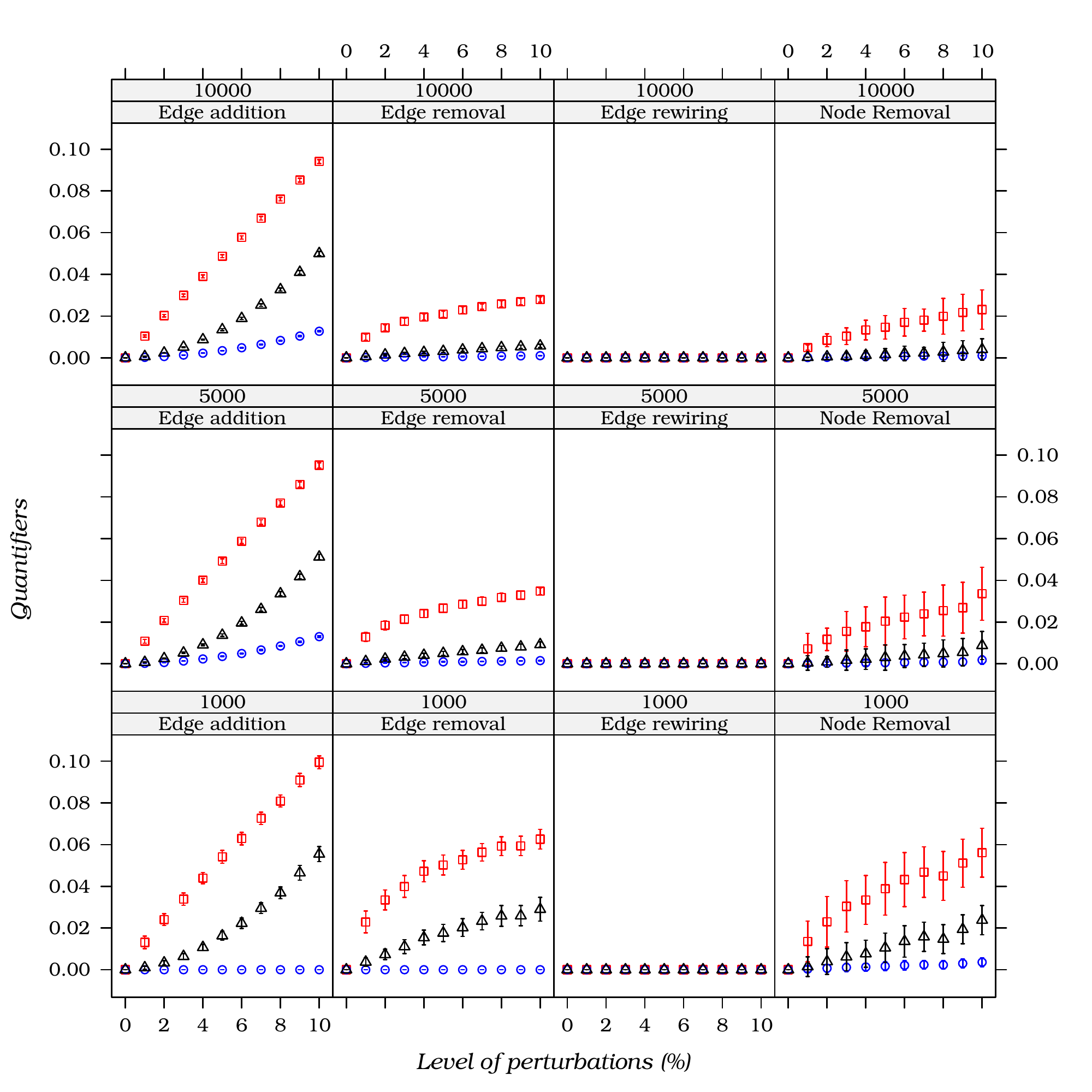}\label{fig:D_BA_k}}
	\caption{Hellinger distance ($\square$), Jensen-Shannon distance ($\circ$) and Kullback-Leibler divergence ($\triangle$) quantifiers for the vertex degree $\kappa_i$.}
	\label{fig:D_k}
\end{center}
\end{sidewaysfigure}

\begin{sidewaysfigure}[htb]
\begin{center}
	\subfigure[Random model]{
	\includegraphics[width=.45\textwidth]{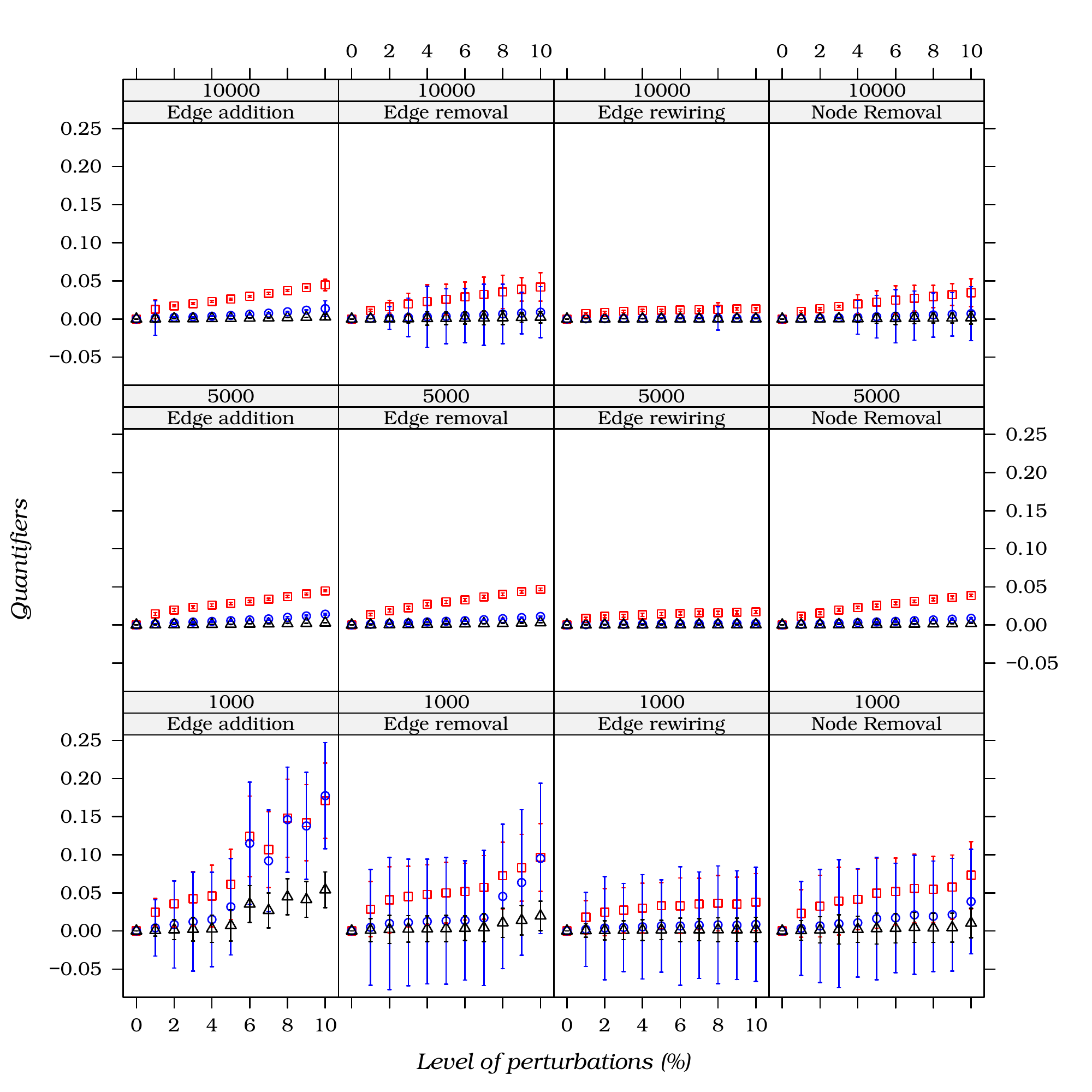}\label{fig:D_ER_B}}
	\subfigure[Small-world Model]{
	\includegraphics[width=.45\textwidth]{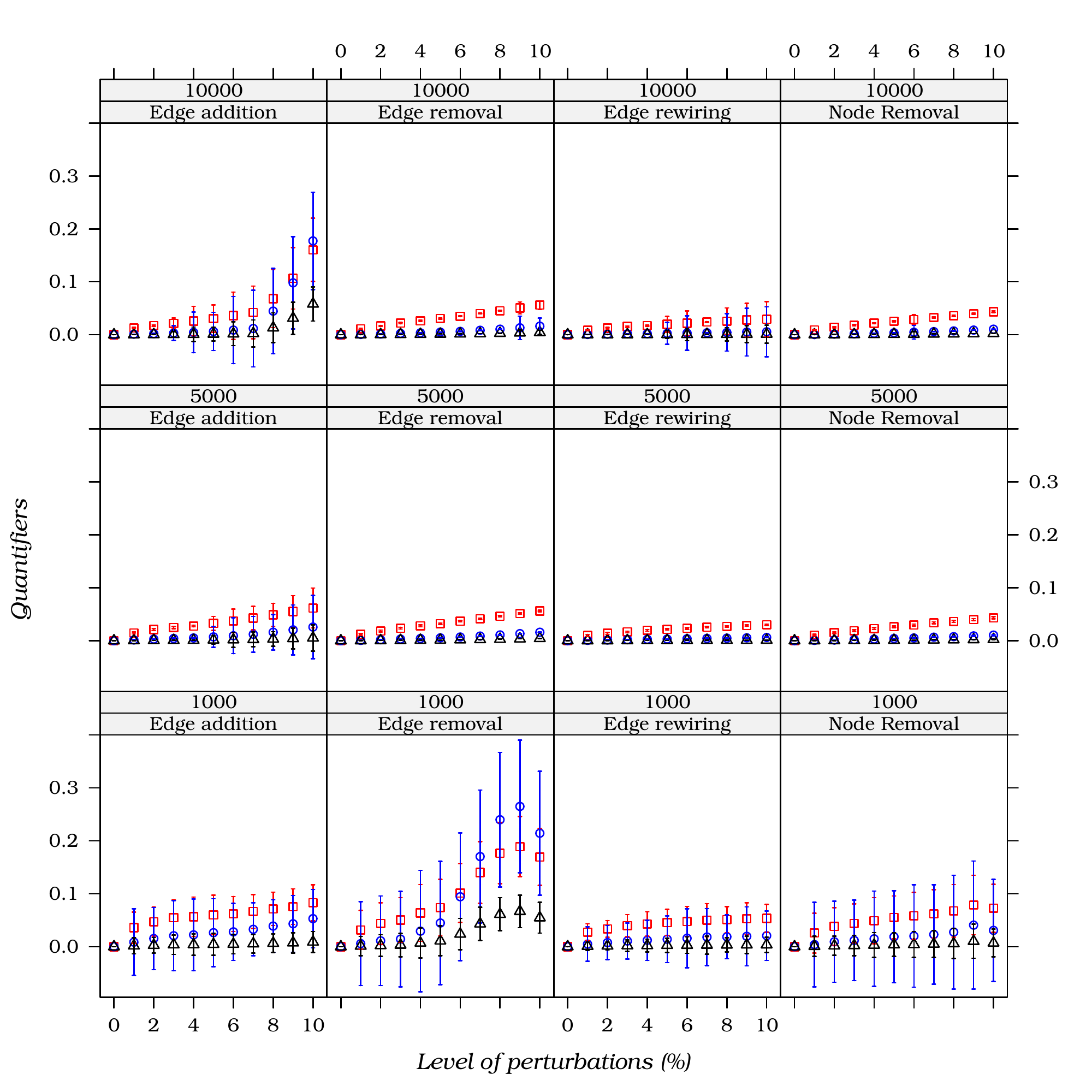}\label{fig:D_WS_B}}
	\subfigure[Scale-free model]{
	\includegraphics[width=.45\textwidth]{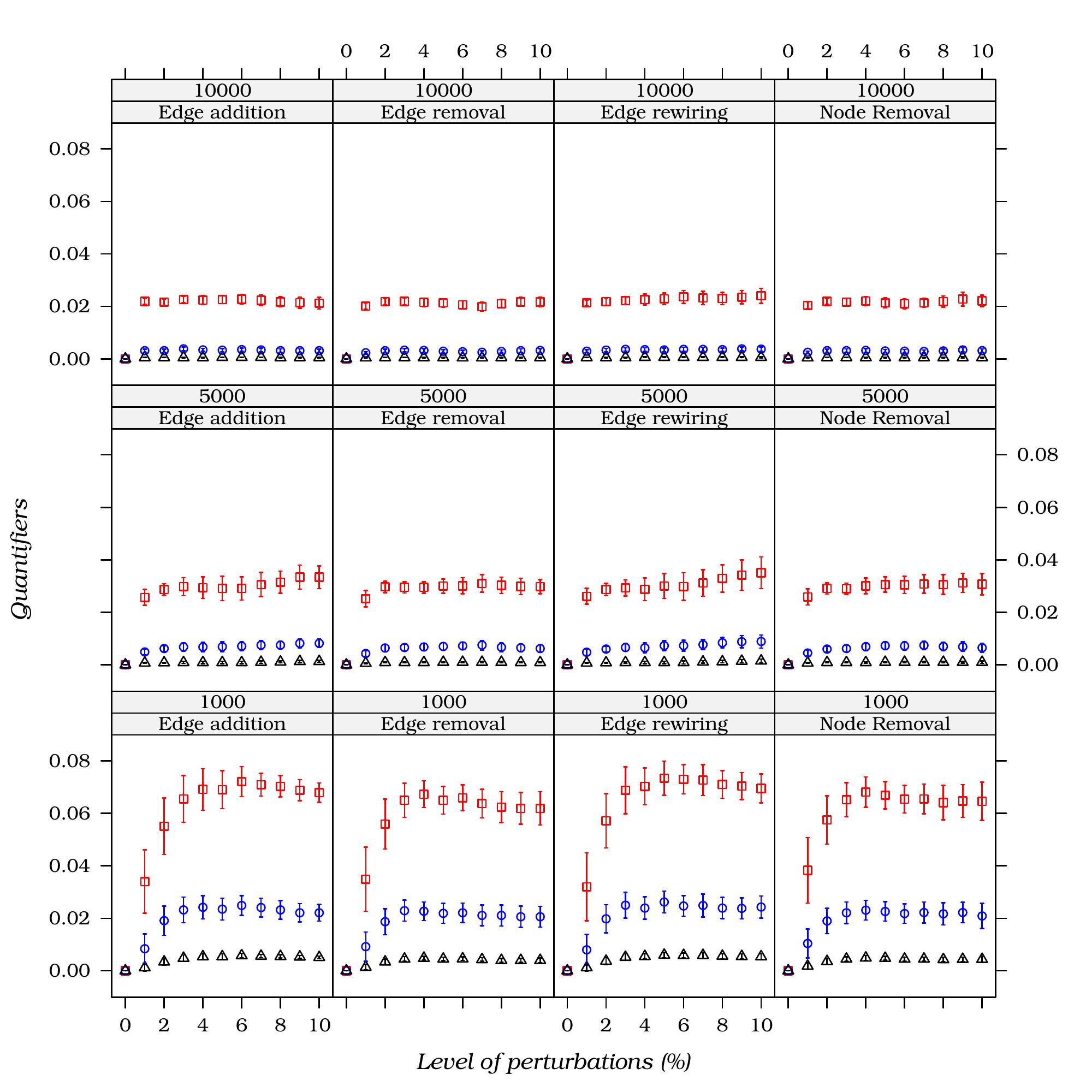}\label{fig:D_BA_B}}
	\caption{Hellinger distance ($\square$), Jensen-Shannon distance ($\circ$) and Kullback-Leibler divergence ($\triangle$) quantifiers for the betweenness centrality $B_i$.}
	\label{fig:D_B}
\end{center}
\end{sidewaysfigure}

Four measures were analyzed with stochastic quantifiers. 
The results are summarized in Table~\ref{tab:c} (clustering coefficient $C_i$), Table~\ref{tab:l} and Figure~\ref{fig:D_l} (shortest paths length $\ell_{(i,j)}$), Table~\ref{tab:kappa} and Figure~\ref{fig:D_k} (vertex degree $\kappa_i$), Table~\ref{tab:b} and Figure~\ref{fig:D_B} (betweenness centrality $B_i$).
The Tables show the maximum values of the quantifiers for $N=10000$ the level perturbation of $10\%$. 

The results for the cluster coefficient $C_i$ are not shown here for brevity.
This measure did not exhibit changes with respect to edge perturbations and to node removal in scale-free model, thus, it can be considered robust or insensitive in these cases. 
This measure, when applied to random and small-world models, is slightly sensitive to node removal, being the Hellinger distance the one which varies most: in mean, $0.052$ in the small-world for the most intense removal of nodes, as we can see in Table~\ref{tab:c}.

\begin{table}[hbt]
\caption{Results for the cluster coefficient $C_i$, $N=10000$ the level perturbation of $10\%$ to node removal $\pi_{rv}$ perturbation.}
\centering
\begin{tabular}{@{}c|*4{r}|*4{r}|*4{r}@{}}
\toprule
 & $\mathcal D_{KL}$ & $\mathcal D_{JS}$ & $\mathcal D_{H}$ \\  \hline
$\mathcal G_{ER}$ & $0.002$ & $10^{-4}$ & $0.018$ \\ 
$\mathcal G_{WS}$ & $0.015$ & $0.003$ & $\textbf{0.052}$ \\ 
\bottomrule
\end{tabular}
\label{tab:c}
\end{table}

\begin{table}[hbt]
\caption{Results for the shortest path length $\ell_{(i,j)}$, $N=10000$ the level perturbation of $10\%$.}
\centering
\setlength{\tabcolsep}{3pt}
\begin{tabular}{@{}c|*4{r}|*4{r}|*4{r}@{}}
\toprule
 & \multicolumn{4}{c|}{$\mathcal D_{KL}$} & \multicolumn{4}{c|}{$\mathcal D_{JS}$} & \multicolumn{4}{c}{$\mathcal D_{H}$} \\  \cline{2-13}
 & $\pi_{ae}$ & $\pi_{re}$  & $\pi_{we}$  & $\pi_{rv}$ & $\pi_{ae}$ & $\pi_{re}$  & $\pi_{we}$ & $\pi_{rv}$ & $\pi_{ae}$ & $\pi_{re}$  & $\pi_{we}$ & $\pi_{rv}$ \\ \hline 
$\mathcal G_{ER}$ & $0.090$ & $0.099$ & \multicolumn{1}{c}{$0$} & $0.065$ & $0.002$ & $0.003$ & \multicolumn{1}{c}{$0$} & $0.017$ & $0.118$ & $0.136$ & $0.001$ & $0.112$ \\ 
$\mathcal G_{WS}$ & $0.183$ & $0.125$ & $0.013$ & $0.091$ &  $0.005$ & $0.004$ & $10^{-4}$ & $0.026$ &  $0.164$ & $0.155$ & $0.047$ & $0.136$ \\ 
$\mathcal G_{BA}$ & $\textbf{0.807}$ & $0.117$ & $0.198$ & $0.121$ & $0.014$ & $0.003$ & $0.005$ & $0.030$ & $0.291$ & $0.134$ & $0.169$ & $0.140$ \\
\bottomrule
\end{tabular}
\label{tab:l}
\end{table}

\begin{table}[hbt]
\caption{Results for the vertex degree $\kappa_i$, $N=10000$ the level perturbation of $10\%$.}
\centering
\setlength{\tabcolsep}{3pt}
\begin{tabular}{@{}c|*4{r}|*4{r}|*4{r}@{}}
\toprule
 & \multicolumn{4}{c|}{$\mathcal D_{KL}$} & \multicolumn{4}{c|}{$\mathcal D_{JS}$} & \multicolumn{4}{c}{$\mathcal D_{H}$} \\  \cline{2-13}
 & $\pi_{ae}$ & $\pi_{re}$  & $\pi_{we}$ & $\pi_{rv}$ & $\pi_{ae}$ & $\pi_{re}$  & $\pi_{we}$ & $\pi_{rv}$ & $\pi_{ae}$ & $\pi_{re}$  & $\pi_{we}$ & $\pi_{rv}$\\ \hline 
$\mathcal G_{ER}$ &  $0.047$ & $0.048$ & \multicolumn{1}{c}{$0$} & $0.047$ &  $0.011$ & $0.011$ & \multicolumn{1}{c}{$0$} & $0.011$ &  $0.090$ & $0.090$ & \multicolumn{1}{c}{$0$} & $0.090$ \\ 
$\mathcal G_{WS}$ &  $0.082$ & $0.087$ & \multicolumn{1}{c}{$0$} & $0.083$ & $0.020$ & $0.021$ & \multicolumn{1}{c}{$0$} & $0.021$ & $0.120$ & $\textbf{0.123}$ & \multicolumn{1}{c}{$0$} & $0.122$ \\ 
$\mathcal G_{BA}$ &  $0.050$ & $0.005$ & \multicolumn{1}{c}{$0$} & $0.004$ &  $0.012$ & $10^{-3}$ & \multicolumn{1}{c}{$0$} & $0.096$ & $0.094$ & $0.027$ & \multicolumn{1}{c}{$0$} & $0.023$ \\
\bottomrule
\end{tabular}
\label{tab:kappa}
\end{table}

\begin{table}[hbt]
\caption{Results for the betweenness centrality $B_i$, $N=10000$ the level perturbation of $10\%$.}
\centering
\setlength{\tabcolsep}{3pt}
\begin{tabular}{@{}c|*4{r}|*4{r}|*4{r}@{}}
\toprule
 & \multicolumn{4}{c|}{$\mathcal D_{KL}$} & \multicolumn{4}{c|}{$\mathcal D_{JS}$} & \multicolumn{4}{c}{$\mathcal D_{H}$} \\  \cline{2-13}
 & $\pi_{ae}$ & $\pi_{re}$  & $\pi_{we}$ & $\pi_{rv}$ & $\pi_{ae}$ & $\pi_{re}$  & $\pi_{we}$ & $\pi_{rv}$ & $\pi_{ae}$ & $\pi_{re}$  & $\pi_{we}$ & $\pi_{rv}$\\ \hline 
$\mathcal G_{ER}$ & $0.002$ & $0.002$ & $10^{-4}$ & $0.001$ & $0.003$ & $0.008$ & $0.001$ & $0.007$ &  $0.044$& $ 0.041$ & $0.013$ & $0.034$ \\ 
$\mathcal G_{WS}$ & $0.008$ & $0.004$ & $0.001$ & $0.002$ & $\textbf{0.177}$ & $0.016$ & $0.005$ & $0.010$ & $0.160$ & $0.056$  & $0.029$ & $0.043$ \\ 
$\mathcal G_{BA}$ & $10^{-4}$ & $10^{-4}$ & $10^{-4}$ & $10^{-4}$ & $0.003$ & $0.003$ & $0.003$ & $0.003$ & $0.021$ & $0.021$ & $0.024$ & $0.022$\\
\bottomrule
\end{tabular}
\label{tab:b}
\end{table}

Each of these figures presents three blocks of plots. 
Blocks labeled ``(a)'' relate the results observed in the random model, blocks ``(b)'' correspond to the small-world model, and blocks ``(c)'' show the results obtained in the scale-free model.
Each block presents a grid of $3 \times 4$ plots. 
The lines correspond to the three graph sizes $1000$, $5000$ and $10000$ (bottom to top rows), while the columns are the four types of perturbations: edge addition, edge removal, edge rewiring and node removal (left to right columns).

The figures present the variation of the three quantifiers with respect to the level of perturbation:
the Hellinger distance $\mathcal{D}_{H}$ as squares ``\scalebox{0.6}{$\square$}'',
the Jensen-Shannon distance $\mathcal{D}_{JS}$ as circles ``\scalebox{1}{$\circ$}'', and
the Kullback-Leibler divergence $\mathcal{D}_{KL}$ as triangles ``\scalebox{0.6}{$\triangle$}''.
The main results are discussed in the following.

As presented in Figure~\ref{fig:D_l}, the \textit{shortest paths} ($\ell_{(i,j)}$) exhibit the same behavior in all models. 

\begin{figure}[htb]
\begin{center}
	\subfigure[Initial network]{\label{fig:ba_graph_20}  
	\includegraphics[width=.3\textwidth]{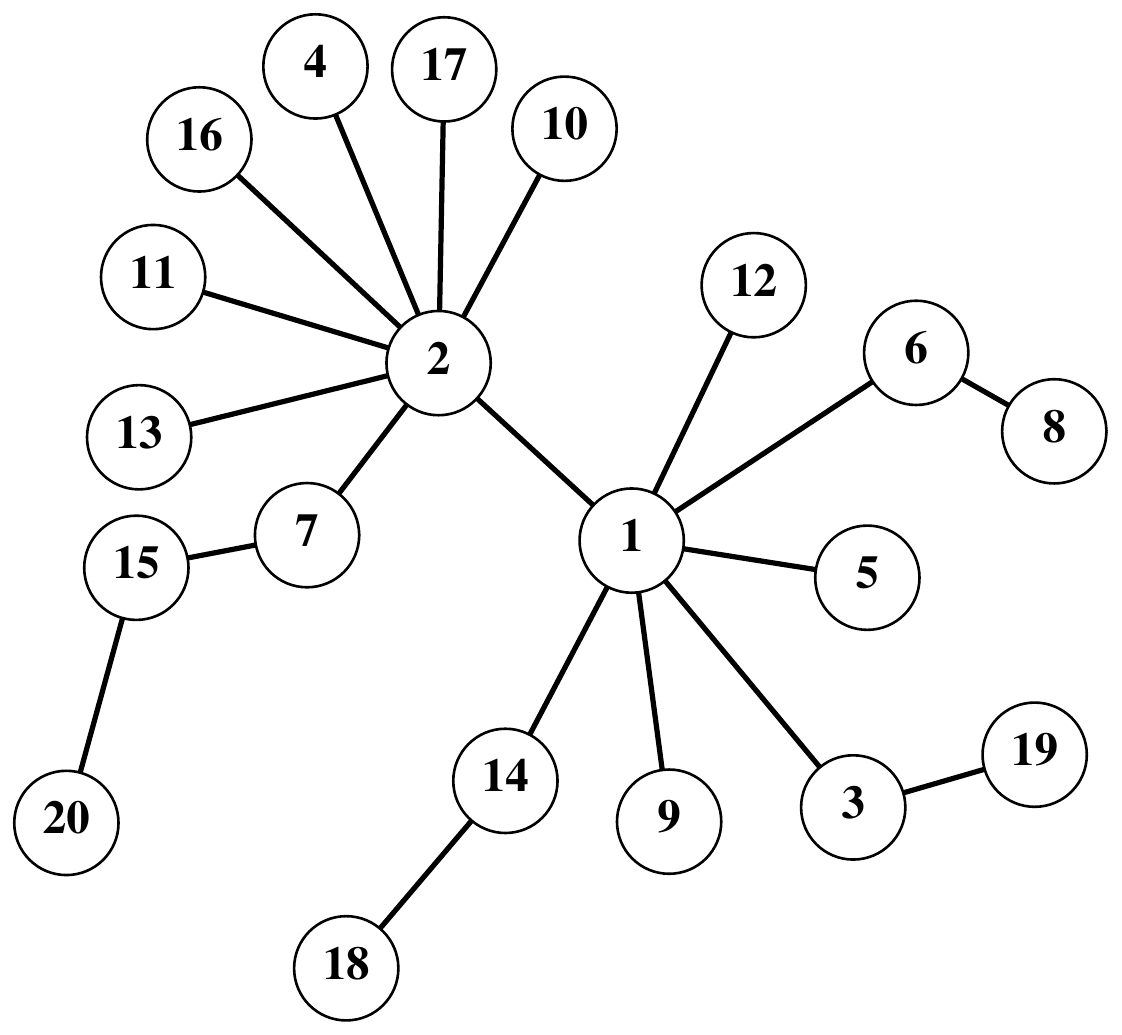}} \hspace{2cm} 
	\subfigure[Network of Fig.~\ref{fig:ba_graph_20} with an edge $e=(18,20)$ added] {\label{fig:ba_graph_20_add_edge}\includegraphics[width=.3\textwidth]{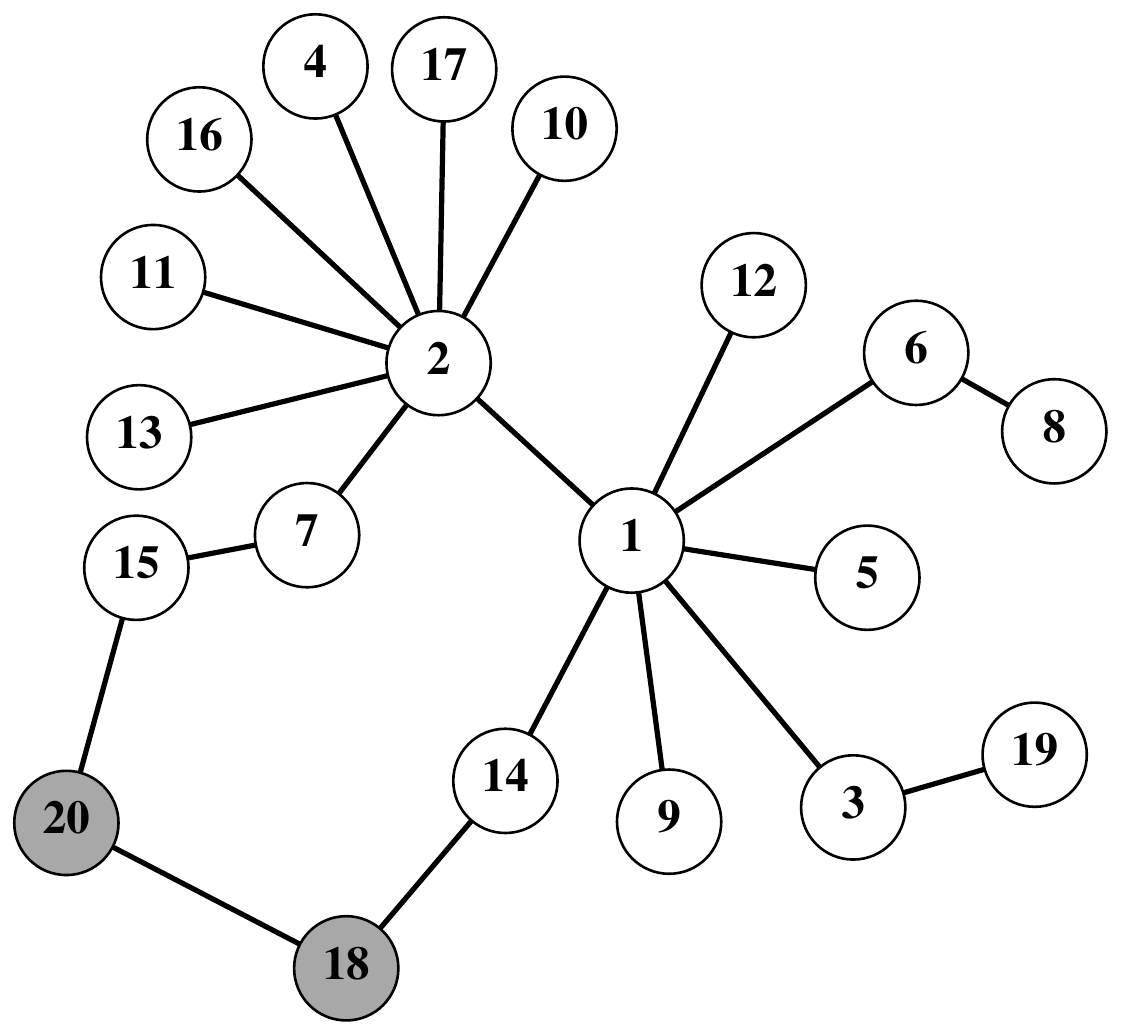}} 
	\subfigure[Network of Fig.~\ref{fig:ba_graph_20} with an edge $e=(1,2)$ deleted]{\label{fig:ba_graph_20_delete_edge}\includegraphics[width=.3\textwidth]{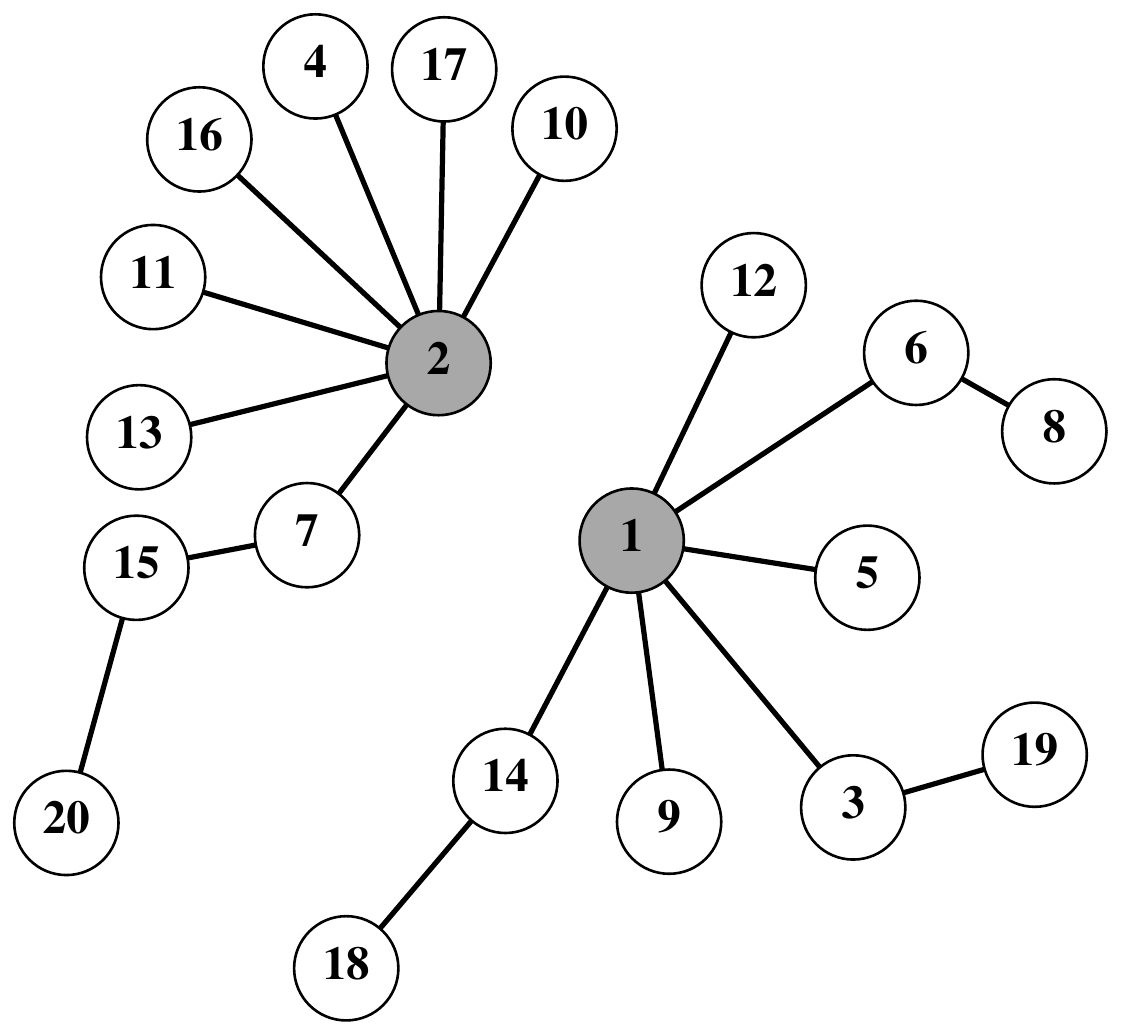}} 
	\caption{Barab\'asi-Albert scale-free network with $N=20$.}
	\label{fig:ba_net_n_20}
\end{center}
\end{figure}

All perturbations increase all the quantifiers, i.e., they render progressively different graphs from the original one as the level of perturbation increases.
The Scale-free model is more sensitive to edge addition than the others, as observed comparing the fist column of Fig.~\ref{fig:D_BA_l} with Figs.~\ref{fig:D_ER_l} and~\ref{fig:D_WS_l}. 
The main feature of these networks is the presence of hubs, which makes them more sensitive to these perturbations; the shortest path length alters radically whenever a hub is added or removed.
Additionally, the removal of some edges makes the network disconnected. 
Figure~\ref{fig:ba_graph_20} shows a example of scale-free network with $N = 20$, where the nodes $1$ and $2$ are hubs.

The shortest path length between nodes $18$ and $20$ on the initial network~\ref{fig:ba_graph_20} is $\ell_{(18,20)} = 6$, but if we add an edge between these vertices, then $\ell_{(18,20)} = 1$, as we see in Figure~\ref{fig:ba_graph_20_add_edge}.
On the other hand, the removal edges in the graph can make the network disconnected, a situation shown in Figu\-re~\ref{fig:ba_graph_20_delete_edge}.
In this case, only the connected vertices are considered for the calculation of $\ell_{(i,j)}$, so it is not subjected to strong changes.
Edge removal in scale-free model introduces a slow growth in quantifiers, cf. second column of Figure~\ref{fig:D_BA_l}.
Edge rewiring does not introduce any noticeable difference to the random model, see the last column of Figure~\ref{fig:D_ER_l}.
This last perturbation introduce a slow growth in the quantifiers of small-world ans scale-free models, see the last column of the figures~\ref{fig:D_WS_l} and~\ref{fig:D_BA_l}, respectively.
Node removal leads to changes in the shortest paths in all network models.

Among the stochastic quantifiers, the most sensitive to the three types of perturbation respecting shortest paths is the Hellinger distance $\mathcal{D}_{H}$, followed by the Kullback-Leibler divergence $\mathcal{D}_{KL}$. 
The Jensen-Shannon distance $\mathcal{D}_{JS}$ is the quantifier with the least variation when edge perturbations are applied, in particular for graphs with $1000$ nodes, it does not exhibit any noticeable change; with $5000$ nodes and edge removal, the Jensen-Shannon $\mathcal{D}_{JS}$ is also flat. 
In scale-free model, among the quantifiers, the most sensitive quantifier to the edge addition is the Kullback-Leibler divergence $\mathcal{D}_{KL}$, see the first column of the block~\ref{fig:D_BA_l}.

\textit{Vertex degree} ($\kappa_i$), as presented in Figure~\ref{fig:D_k}, exhibits the same behavior in all models. 
Edge addition, edge removal and node removal increase all the quantifiers, i.e., they render progressively different graphs from the original one as the level of perturbation increases. 
Edge rewiring does not introduce any noticeable difference, cf. the last column of the three blocks.
Observe that, the confidence intervals are larger for node removal in small-world ($1000$ vertices) and scale-free ($1000$ and $5000$ vertices) models, however one can check an asymptotic behavior.

Among the stochastic quantifiers, the most sensitive to the two types of perturbation which induce changes is the Hellinger distance $\mathcal{D}_{H}$, followed by the Kullback-Leibler divergence $\mathcal{D}_{KL}$. 
The Jensen-Shannon distance $\mathcal{D}_{JS}$ is the quantifier with the least variation when edge addition and edge removal are applied, and when the graph has $1000$ nodes it does not exhibit any noticeable change; with $5000$ nodes and edge removal, the Jensen-Shannon $\mathcal{D}_{JS}$ is also flat.

Observe that the confidence intervals are larger for $1000$ vertices and node removal in random and  small-world models.
Although any pairwise comparison of subsequent values would not lead to the rejection of the hypothesis that the values change, the overall behavior of the mean provides enough information to assess its dependence on this property of node removal perturbation.   

As presented in Figure~\ref{fig:D_B}, \textit{betweenness centrality} ($B_i$) exhibits the same behavior in random and small-world models. 
All perturbations increase all the quantifiers, i.e., they render progressively different graphs from the original one as the level of perturbation increases.
Notice that the confidence intervals are large in some cases, as we can see in the first line (top to down) of Figures~\ref{fig:D_ER_B} and~\ref{fig:D_WS_B}, but the results provide enough information to assess the dependence of this measure on the perturbations.

Regarding the scale-free model, both measures vary strongly when small perturbations are applied, but the change tends to stabilize, i.e., it saturates, soon after, as we see in Figure~\ref{fig:D_BA_B}.

The most sensitive quantifier to the perturbations (for $B_i$), is the Hellinger distance $\mathcal{D}_{H}$, followed by the Kullback-Leibler divergence $\mathcal{D}_{KL}$. 
The Jensen-Shannon distance $\mathcal{D}_{JS}$ is the quantifier with the least variation.

In summary, among all perturbations, the edge rewiring causes smaller variations in the measures, in particular the vertex degree ($\kappa_i$) and the cluster coefficient ($C_i$) did not change with this perturbation.
Edge addition, edge removal and node removal affect directly the behavior of measures $\ell_{(i,j)}$, $\kappa_i$ and $B_i$.

The shortest paths ($\ell_{(i,j)}$) changed in all network models, but it is most sensitive to perturbations applied the scale-free model.
In particular, $\mathcal{D}_{H}$ had the highest values, i.e., this distance captured more accurately the changes in the network, as we see in the first column of Figure~\ref{fig:D_BA_l}: mean of $0.801$ for the most intense removal of nodes (Table~\ref{tab:l}).

The values of $\mathcal{D}_{KL}$ and $\mathcal{D}_{H}$ indicate that the networks are sensitive to edge addition, edge removal and node removal.
The cluster coefficient ($C_i$) is not sensitive to edge perturbations and it is slightly sensitive to node removal.

The maximum value of $\mathcal{D}_{KL}$ was $0.807$ (mean). 
In this case, an edge was added between hubs and the shortest paths decreased (the same case of the example of Figure~\ref{fig:ba_net_n_20}).
The  Jensen-Shannon distance $\mathcal{D}_{JS}$ had the lowest values for the measures $\kappa_i$, $\ell_{(i,j)}$ and $C_i$, i.e. these are not sensitive to the perturbations applied to the networks under assessment.
The Jensen-Shannon distance did not vary with $N = 1000$ vertices and showed a slight variation with $N = \{5000,10000\}$ vertices; as we see in Fig.~\ref{fig:D_ER_l}, the maximum $\mathcal{D}_{JS}$ is $0.002$
for the most intense addition of edges (Table~\ref{tab:l}). 

The degree distribution is known for the models here considered, see equations~\eqref{eq:pk_er}, \eqref{eq:pk_ws} and~\eqref{eq:pk_sf} for the random, small-world and scale-free models, respectively.
It is, therefore, possible to apply hypothesis tests to check whether it changed or not and, if it changed, if the distribution was altered.
To perform the tests we use networks with $N=1000$. 

Two hypotheses were verified: 
\begin{enumerate}
\item the degree distribution of the networks was preserved, i.e., $H_0:p_{g_j}(\kappa) = p_{\eta_{ij}}(\kappa)$ for every $\kappa$, and 
\item the degree was preserved, even if the distribution changed (first order property).
\end{enumerate}
We used the Kolmogorov-Smirnov test for the former and the Student's $t$ test for the latter.

\begin{sidewaysfigure}[htb]
	\centering
	\subfigure[Kolmogorov-Smirnov test]{
	\includegraphics[width=.46\textwidth]{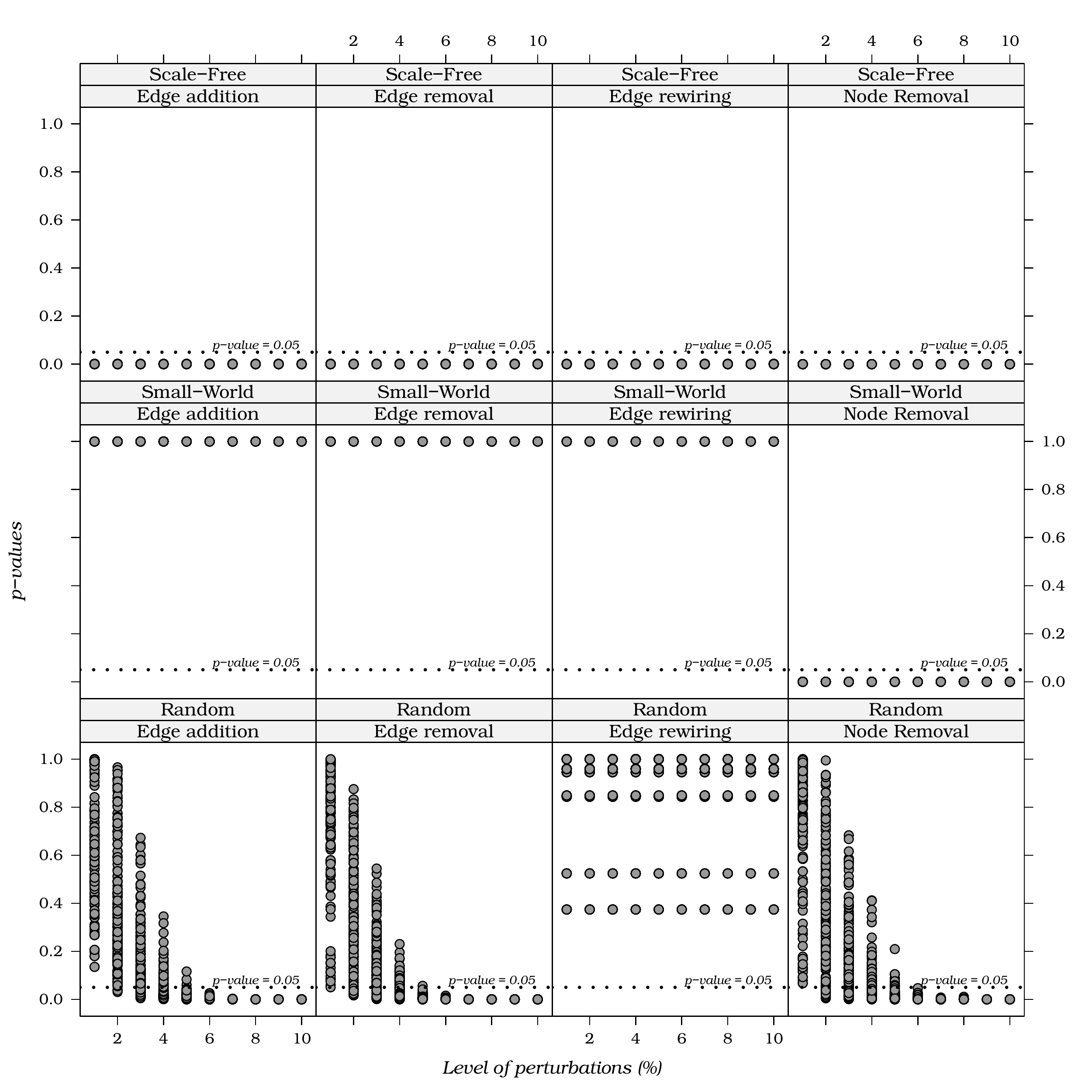}\label{fig:ks-test}}
	  \subfigure[Student's $t$ test]{
	  \includegraphics[width=.46\textwidth]{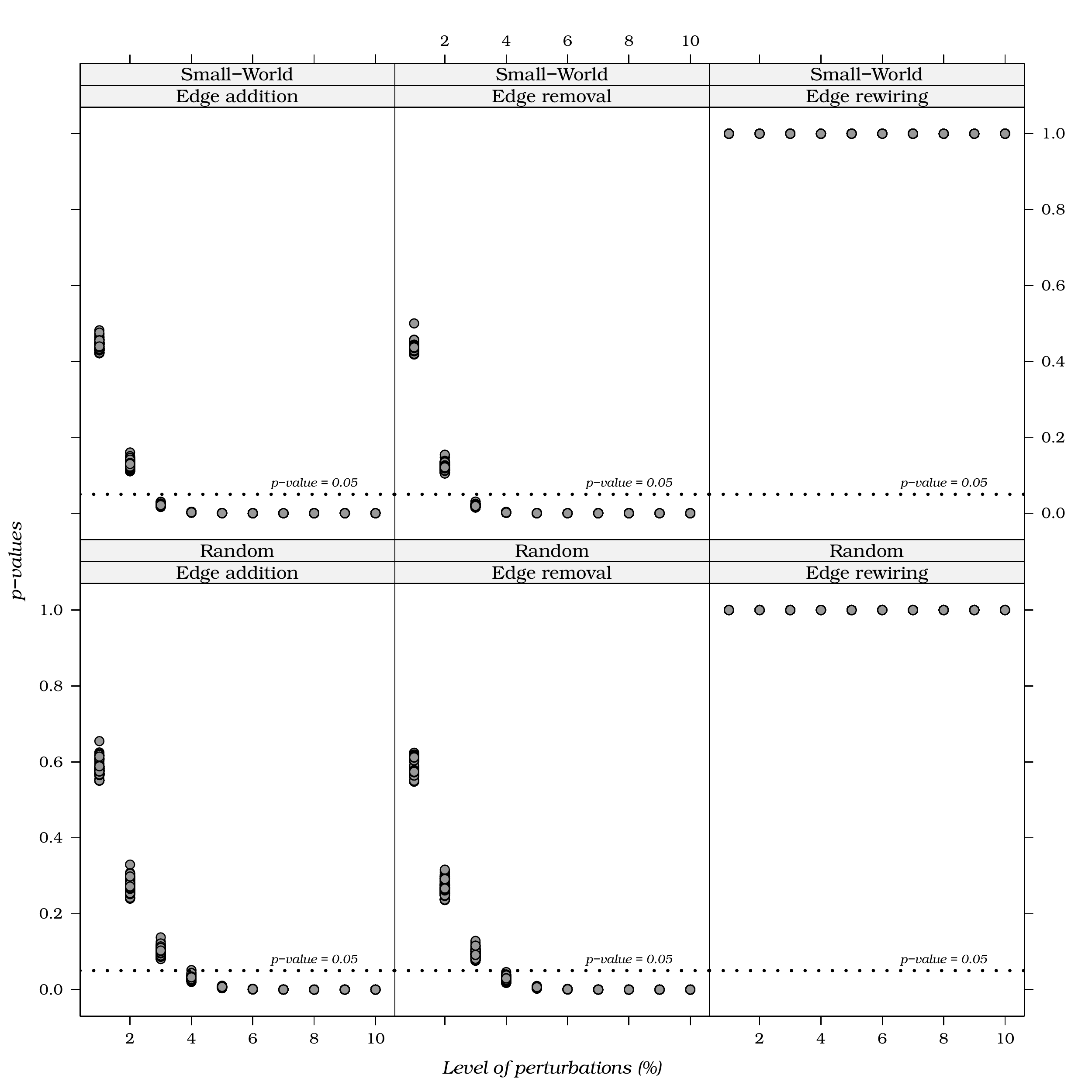}\label{fig:t-test}}	  
	  \caption{$P$-values for the hypotheses tests for networks with $N=1000$.
	  Dashed line are at $0.05$.}
\end{sidewaysfigure}

Figure~\ref{fig:ks-test} shows the results in the small-world, scale-free and random models (top to bottom rows) and the four types of perturbations: edge addition, edge removal, edge rewiring and node removal (left to right columns), while Table~\ref{tab:count_pvalues_ks-test} presents the percentage of $p$-values which did not reject the Kolmogorov-Smirnov test at the $5\%$ level of significance.

The first hypothesis $H_0$ was rejected in both the random and scale-free models after the perturbations were applied. 
In the first model, when the level of perturbations increases, the $p$-values are close to zero, so $H_0$ was rejected whenever the perturbation is greater than $4\%$, see Table~\ref{tab:count_pvalues_ks-test}.
In the scale-free model, all $p$-values are smaller than $0.05$ for all levels of perturbations.

In particular, edge rewiring does not lead to the rejection of $H_0$ in the random model, as we observe in Figure~\ref{fig:ks-test} (first line from down to top and third column from left to right).
$H_0$ was not rejected in the three types of edge perturbations applied to the small-world model, since the all $p$-values are greater than $0.05$ but, otherwise, the hypothesis is rejected for node removal.

We may say that whenever $H_0$ is rejected, the perturbations has led to a breakdown of the degree distribution of the network.
 
\begin{table}[htb]
\setlength{\tabcolsep}{2pt}
	\caption{Percentage of $p$-values $\geq$ $0.05$ for the Kolmogorov-Smirnov test for networks with $N=1000$ and levels of perturbation $1\%$, $4\%$ and $10\%$. A total of $200$ samples were tested.}
	\label{tab:count_pvalues_ks-test}	
	 \centering
	\begin{tabular}{@{}c|ccc|ccc|ccc|ccc@{}}
	\toprule
	 & \multicolumn{3}{c|}{$\pi_{ae}$} &  \multicolumn{3}{c|}{$\pi_{re}$}  & \multicolumn{3}{c|}{$\pi_{we}$} & \multicolumn{3}{c}{$\pi_{rv}$} \\ \cline{2-13}
	& $1\%$ & $4\%$ & $10\%$ & $1\%$ & $4\%$ & $10\%$ & $1\%$ & $4\%$ & $10\%$ & $1\%$ & $4\%$ & $10\%$ \\  \midrule
	$\mathcal{G}_{BA}$  & $0$ & $0$ & $0$ & $0$ & $0$ & $0$ & $0$ & $0$  & $0$  & $0$ & $0$ & $0$ \\ 
	$\mathcal{G}_{WS}$ & $100$ & $100$ & $100$ & $100$ & $100$ & $100$ & $100$ & $100$ & $100$ & $0$ & $0$ & $0$ \\
	$\mathcal{G}_{ER}$ & $100$ & $30$ & $0$ & $100$ & $25$ & $0$ & $100$ & $100$ & $100$ & $100$ & $35$ & $0$ \\ 
	\bottomrule
	\end{tabular}
\end{table}

The Student's $t$ test checks if the degree was preserved after the perturbations.
As the Kolmogorov-Smirnov test showed that the degree distribution was preserved only for small-world and random model, we apply the Student's $t$ only to these models and to edge perturbations (where the number of nodes is the same).
Figure~\ref{fig:t-test} shows the results in the small-world and random networks (top to bottom rows) and the three types of perturbation: edge addition, edge removal and edge rewiring (left to right columns), while Table~\ref{tab:count_pvalues_t-test} presents the percentage of rejected situations.

In the small-world and random models, as the level of edge addition and removal increases the $p$-values are close to zero, the degrees was not preserved in perturbations greater than $3\%$ for the first and $4\%$ for the last. 
As expected, edge rewiring did not change the degree (all $p$-values greater than $0.05$).

\begin{table}[htb]
\setlength{\tabcolsep}{2pt}
\caption{Percentage of $p$-values $\geq$ $0.05$ for the Student's $t$ test for networks with $N=1000$ and levels of perturbation $1\%$, $4\%$ and $10\%$. 
A total of $200$ samples were tested.}
\label{tab:count_pvalues_t-test}	
  \centering
    \begin{tabular}{@{}c|ccc|ccc|ccc@{}}
    \toprule
    & \multicolumn{3}{c|}{$\pi_{ae}$} &  \multicolumn{3}{c|}{$\pi_{re}$}  & \multicolumn{3}{c}{$\pi_{we}$} \\ \cline{2-10}
    & $1\%$ & $4\%$ & $10\%$ & $1\%$ & $4\%$ & $10\%$ & $1\%$ & $4\%$ & $10\%$ \\  \midrule
    $\mathcal{G}_{WS}$ & $100$ & $0$ & $0$ & $100$ & $0$ & $0$ & $100$ & $100$ & $100$  \\
    $\mathcal{G}_{ER}$ & $100$ & $0\%$ & $0$ & $100$ & $0$ & $0$ & $100$ & $100$ & $100$ \\ 
    \bottomrule
    \end{tabular}
\end{table}

\subsection{Real-world Networks}

Two real networks were analyzed with stochastic quantifiers: a metabolic and a collaboration network. 

The \textit{metabolic network} was evaluated with respect to edge addition, edge removal and node removal.
For brevity, we only show the results for the $10\%$ level of perturbation in Table~\ref{tab:metabolic}.
These results show that the variation of quantifiers is very small, being the Hellinger and Jensen-Shannon distances the ones which varies most and least, respectively. 

Measures $\ell_{(i,j)}$ and $C_i$ exhibit the same behavior: perturbations progressively increase the values of the quantifiers, and are more sensitive to edge addition; $C_i$ presents the highest value to edge addition.

Measures $\kappa_i$ and $B_i$ behave alike. 
Edge addition increases the quantifiers at low levels of perturbation ($1\%$) and then they tend to become constant as the level of perturbation increases; this perturbation, thus, induces a kind of saturation in the measures.
Edge removal and node removal increase progressively but slightly the values of the quantifiers.

\begin{table}[hbt]
\caption{Results for measures in metabolic networks, the level perturbation of $10\%$ to node perturbations.}
\centering
\begin{tabular}{@{}c|*3{r}|*3{r}|*3{r}@{}}
\toprule
 & \multicolumn{3}{c|}{$\mathcal D_{KL}$} & \multicolumn{3}{c|}{$\mathcal D_{JS}$} & \multicolumn{3}{c}{$\mathcal D_{H}$} \\  \cline{2-10}
& $\pi_{ae}$ & $\pi_{re}$ & $\pi_{rv}$ & $\pi_{ae}$ & $\pi_{re}$  & $\pi_{rv}$ & $\pi_{ae}$ & $\pi_{re}$  & $\pi_{rv}$\\ \hline  

$\ell_{(i,j)}$ & $0.025$  & $0.002$ & $0.008$ & $0.005$ & $10^{-4}$ & $0.002$  & $\textbf{0.062}$ & $0.022$ & $0.038$ \\ 

$\kappa_i$ & $0.014$  & $0.023$ & $0.019$ & $0.003$ & $0.004$ &  $0.004$ & $0.053$ & $0.063$ & $\textbf{0.066}$ \\ 

$C_i$ & $\textbf{0.174}$ & $0.004$ & $0.018$ &$0.038$ & $0.001$& $0.004$ & $0.166$ & $0.027$& $0.056$ \\ 

$B_i$ & $0.023$  & $0.005$ & $0.025$ & $0.003$ & $10^{-4}$ & $0.003$ & $0.055$ & $0.020$ & $\textbf{0.058}$ \\ 
\bottomrule
\end{tabular}
\label{tab:metabolic} 
\end{table}

The \textit{collaboration network} was evaluated with respect to edge addition, edge removal and node removal.
We observed that the intensity the perturbations did not affect the behavior of the measures degree ($\kappa$), betweenness ($b$) and shortest paths ($\ell$), the values of the quantifiers are very small as can be seen in Table~\ref{tab:collaboration}.
The cluster coefficient varies when the network was subjected to edge perturbations, the values of quantifiers increase progressively with increasing levels of perturbations. 
This behavior is due to the fact that these networks are highly clustered.

\begin{table}[hbt]
\caption{Results for measures in collaboration networks, the level perturbation of $10\%$ to perturbations applied.}
\centering
\begin{tabular}{@{}c|*3{r}|*3{r}|*3{r}@{}}
\toprule
 & \multicolumn{3}{c|}{$\mathcal D_{KL}$} & \multicolumn{3}{c|}{$\mathcal D_{JS}$} & \multicolumn{3}{c}{$\mathcal D_{H}$} \\  \cline{2-10}
& $\pi_{ae}$ & $\pi_{re}$ & $\pi_{rv}$ & $\pi_{ae}$ & $\pi_{re}$  & $\pi_{rv}$ & $\pi_{ae}$ & $\pi_{re}$  & $\pi_{rv}$\\ \hline  

$\ell_{(i,j)}$ & $\textbf{0.176}$ & $0.007$ & $0.002$ & $0.033$ & $0.001$ & $10^{-4}$ & $0.156$ & $0.036$ & $0.022$ \\ 

$\kappa_i$ & $0.014$ & $0.018$ & $0.019$ & $0.003$ & $0.004$ & $0.004$ & $0.053$ & $0.063$ & $\textbf{0.066}$\\ 

$C_i$ & $0.159$ & $0.145$ & $0.009$ & $0.043$ & $0.041$  & $0.002$ & $\textbf{0.173}$ & $0.170$ & $0.040$\\ 

$B_i$ & $0.016$ & $0.002$ & $0.016$ & $0.002$ & $10^{-4}$ & $0.002$ & $0.046$ & $0.015$ & $\textbf{0.048}$\\ 
\bottomrule
\end{tabular}
\label{tab:collaboration} 
\end{table}

\section{Conclusions} \label{sec:conclusion}

The analysis of the variability of measures in complex networks provides important information.
It gives an insight of the behavior of the network when it is perturbed, and it helps us in the design of appropriate solutions for many applications.

In this paper, we used the Hellinger and Jensen-Shannon distances, and the normalized Kullback-Leibler divergence to compare different states of networks: samples from the random, small-world, scale-free, metabolic and collaboration networks. 

We performed four types of perturbations: addition, removal, rewiring of edges and removal nodes. 
We analyzed how path length, vertex degree, cluster coefficient and betweenness centrali\-ty change with respect to different levels of each perturbation. 

The results showed that the clustering coefficient is not sensitive to these perturbations in theoretical models, but is very sensitive in highly clustered networks.

The other measures are sensitive to these changes and, in most situations, they alter their values accordingly to the intensity of the perturbation.
Our analysis promotes the identification of the relationship between the strength of the perturbations and the change of the shortest path length.

We applied hypothesis tests which allowed the identification of which perturbations lead to breakdowns in the degree distribution of the networks. 

The use of quantifiers that involve logarithms or ratios may not be a good choice for this kind of characterization, because the occurrence of zeros leads to numerical problems and, possibly, to incorrect interpretation of network changes.

\section*{Acknowledgement}
This work is partially supported by the Brazilian National Council for Scientific and Technological Development (CNPq).
\bibliographystyle{model1a-num-names}
\bibliography{article}
\end{document}